\documentclass[a4paper,fleqn,usenatbib,useAMS]{mnras}

\usepackage{graphicx}	
\usepackage{amsmath}	
\usepackage{amssymb}	
\usepackage{multicol}        
\usepackage{bm}		
\usepackage{float}
\usepackage{gensymb}
\usepackage[usenames, dvipsnames]{color}

\usepackage{subfigure}

\usepackage[T1]{fontenc}
\usepackage{ae,aecompl}

\title[Centroid vetting of planet candidates from NGTS]{Centroid vetting of transiting planet candidates from the Next Generation Transit Survey}

\author[M.~N.~G{\"u}nther]{
\parbox{\textwidth}{
Maximilian~N.~G{\"u}nther$^{c}$\thanks{E-mail: \href{mg719@cam.ac.uk}{mg719@cam.ac.uk}}, 
Didier~Queloz$^{c}$, 
Edward~Gillen$^{c}$,
James~McCormac$^{w}$,
Daniel~Bayliss$^{g}$,
Francois~Bouchy$^{g}$,
Simon.~R.~Walker$^{w}$,
Richard~G.~West$^{w}$,
Philipp~Eigm\"uller$^{d}$,
Alexis~M.~S.~Smith$^{d}$,
David~J.~Armstrong$^{w}$,
Matthew~Burleigh$^{l}$,
Sarah~L.~Casewell$^{l}$,
Alexander~P.~Chaushev$^{l}$,
Michael~R.~Goad$^{l}$,
Andrew~Grange$^{l}$,
James~Jackman$^{w}$,
James~S.~Jenkins$^{uc,ci}$,
Tom~Louden$^{w}$,
Maximiliano~Moyano$^{a}$,
Don~Pollacco$^{w}$,
Katja~Poppenhaeger$^{q}$,
Heike~Rauer$^{d}$,
Liam~Raynard$^{l}$,
Andrew~P.~G.~Thompson$^{q}$,
St\'{e}phane~Udry$^{g}$,
Christopher~A.~Watson$^{q}$,
Peter~J.~Wheatley$^{w}$
}
\\
$^{c}$Astrophysics Group, Cavendish Laboratory, J.J. Thomson Avenue, Cambridge CB3 0HE, UK\\
$^{w}$Dept.\ of Physics, University of Warwick, Gibbet Hill Road, Coventry CV4 7AL, UK\\
$^{d}$Institute of Planetary Research, German Aerospace Center, Rutherfordstrasse 2, 12489 Berlin, Germany\\
$^{q}$Astrophysics Research Centre, School of Mathematics and Physics, Queen's University Belfast, BT7 1NN Belfast, UK\\
$^{l}$Department of Physics and Astronomy, Leicester Institute of Space and Earth Observation, University of Leicester, LE1 7RH, UK\\
$^{g}$Observatoire de Gen{\`e}ve, Universit{\'e} de Gen{\`e}ve, 51 Ch. des Maillettes, 1290 Sauverny, Switzerland\\
$^{uc}$Departamento de Astronom\'ia, Universidad de Chile,
Casilla 36-D, Santiago, Chile\\
$^{ci}$Centro de Astrof\'isica y Tecnolog\'ias Afines (CATA), Casilla 36-D, Santiago, Chile
$^{a}$Instituto de Astronom\'ia, Universidad Cat\'olica del Norte, Antofagasta, Chile.
}

\date{Last updated -; in original form -}

\pubyear{2017}

\begin{document}
\label{firstpage}
\pagerange{\pageref{firstpage}--\pageref{lastpage}}
\maketitle

\begin{abstract}
The Next Generation Transit Survey (\textit{NGTS}), operating in Paranal since 2016, is a wide-field survey to detect Neptunes and super-Earths transiting bright stars, which are suitable for precise radial velocity follow-up and characterisation. Thereby, its sub-mmag photometric precision and ability to identify false positives are crucial. Particularly, variable background objects blended in the photometric aperture frequently mimic Neptune-sized transits and are costly in follow-up time. 
These objects can best be identified with the centroiding technique: if the photometric flux is lost off-centre during an eclipse, the flux centroid shifts towards the centre of the target star. Although this method has successfully been employed by the Kepler mission, it has previously not been implemented from the ground.
We present a fully-automated centroid vetting algorithm developed for \textit{NGTS}, enabled by our high-precision auto-guiding. Our method allows detecting centroid shifts with an average precision of $0.75$~milli-pixel, and down to $0.25$~milli-pixel for specific targets, for a pixel size of $4.97$~arcsec. 
The algorithm is now part of the \textit{NGTS} candidate vetting pipeline and automatically employed for all detected signals.
Further, we develop a joint Bayesian fitting model for all photometric and centroid data, allowing to disentangle which object (target or background) is causing the signal, and what its astrophysical parameters are.
We demonstrate our method on two \textit{NGTS} objects of interest.
These achievements make \textit{NGTS} the first ground-based wide-field transit survey ever to successfully apply the centroiding technique for automated candidate vetting, enabling the production of a robust candidate list before follow-up.
\end{abstract}

\begin{keywords}
planets and satellites: detection, eclipses, occultations, surveys, (stars:) binaries: eclipsing
\end{keywords}




\section{Introduction}
\label{s:Introduction}

Transiting exoplanets allow determination of their radii relative to their host star. If the host star is bright enough, the planet mass is accessible through radial velocity monitoring. Together, these allow the planet density and hence bulk composition to be determined, making such planets favourable targets for detailed characterisation of their atmospheric structure and composition.
Gaining this insight is a key factor in the search for habitable planets.
However, given the design of previous surveys, the known transiting planet population is typically faint (V > 13).
The Next Generation Transit Survey (\textit{NGTS}) \citep[Wheatley et al., in prep.,][]{Wheatley2013,Chazelas2012} is a wide-field survey designed to detect Neptune-sized exoplanets orbiting bright host stars. The survey commenced full operation at ESO's Paranal Observatory in early 2016.

\begin{figure}
 \includegraphics[width=\columnwidth]{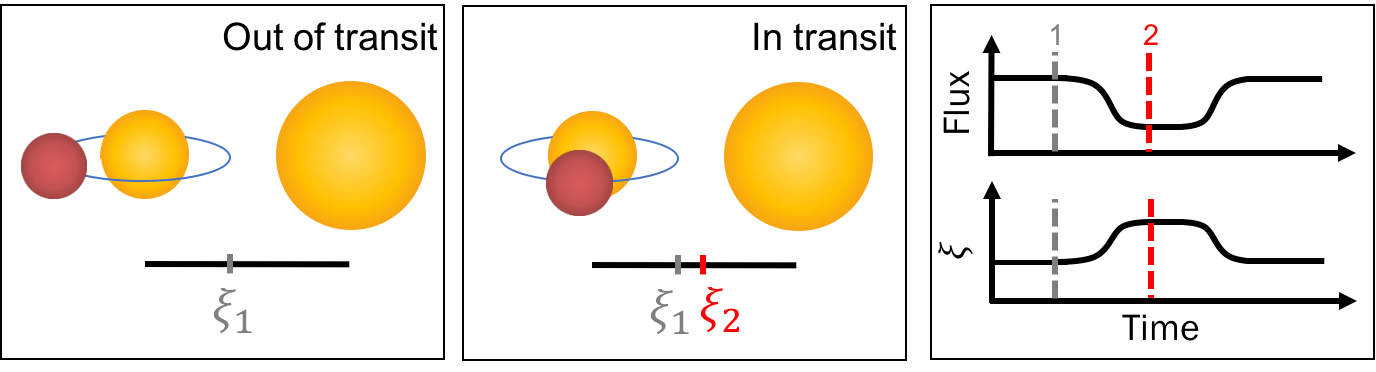}
 \caption{Sketched illustration of a centroid shift $\Delta \xi = \xi_2 - \xi_1$ correlated to a transit-like signal, which is caused by a background eclipsing binary diluted in the aperture of a constant target star. Typically, both systems are photometrically extracted as a single source, and are not visually resolvable in \textit{NGTS} images. Note that if the system can be visually resolved, the direction of the shift indicates which object undergoes the eclipse.}
 \label{fig:sketch}
\end{figure}

In addition to planets, transit-like signals in light curves can be caused by astrophysical false positives, such as eclipsing binaries, which can be incorrectly interpreted as bona fide planetary transits \citep[see e.g.][]{Cameron2012}.
In the case of \textit{NGTS}, we previously showed that the ability to identify false positives, which outnumber the planet yield by an order of magnitude, is expected to be a major factor for the survey's scientific success \citep{Guenther2017}. 
In particular, variable background objects (blended in the photometric aperture) can mimic Neptune-sized transits and are costly in follow-up time (Fig.~\ref{fig:sketch}). 
Such variable background objects can best be identified with the centroiding technique: if the photometric flux is lost off-centre during an eclipse, the flux centroid shifts towards the centre of the target star. Although this method has successfully been employed by the space-based \textit{Kepler} mission (\citealt{Batalha2010} and \citeyear{Batalha2012}, \citealt{Bryson2013}), it has previously not been proven feasible for ground-based surveys.

If a background star lies within the photometric aperture of a target star, the centre of flux in the aperture, $\mathbf{\xi}$, is offset from the true centre of the target (Fig.~\ref{fig:sketch}).
Assuming a symmetric point-spread-function, we introduce the concept of a `photometric centre of mass'. One can directly translate this principle from classical mechanics into the photometric scenario by replacing the term `mass' with `flux', denoted as $F_c(t) = F_c$ for the constant object in the aperture and $F_e(t)$ for the eclipsing object:
\begin{equation}
\label{eq:xi}
\vec{\xi}(t) = \frac{ F_c \ \vec{x}_c + F_e(t) \ \vec{x}_e } { F_c + F_e(t) }.
\end{equation}
The CCD position of the two objects is denoted as $\vec{x}_c$ and $\vec{x}_e$, respectively. Consequently, any change in brightness of one object leads to a shift of the centre of flux in the aperture.
Any centroid shift is hence dependent on $F_e(t)$:
\begin{equation}
\label{eq:Deltaxi}
\Delta \vec{\xi}(t) = \vec{\xi}(t) - \vec{\xi}(t=0)
\end{equation}
A detailed derivation can be found in section~\ref{ss:Analysis of blended systems}.
Note that these equations are for two objects, yet this model allows for an arbitrary number of objects, which are described by their common centre of flux.

In the following, we distinguish four distinct cases. Thereby, we assume that the brighter object in the aperture will be identified as the `target', and denote the fainter objects in the aperture as `blended' or `background objects':
\begin{enumerate}
\item Diluted planet (\textit{dilP}): The target hosts a transiting planet, the background objects are constant (on the respective time-scales and period of the detected signal). The planet transit signal is diluted and the measured depth is decreased. For example, a Hot Jupiter might be miss-identified as a Neptune-sized planet. Note that this means that detecting a correlation between flux and centroid data is not sufficient to disregard a planet candidate.
\item Diluted eclipsing binary (\textit{dilEB}): The target is an eclipsing binary, the background objects are constant. The binary's eclipse signal is diluted and the measured depth is decreased. If the dilution is high and/or the eclipse is shallow, this can mimic a planetary transit.
\item Background planet (\textit{BP}): The target is constant, one of the background objects hosts a transiting planet The transit signal is diluted and the measured depth is decreased. The transit depth would be decreased by $>50\%$, in most cases hindering the detection of the signal.
\item Background eclipsing binary (\textit{BEB}): The target is constant, one of the background objects is an eclipsing binary. The transit signal is diluted and the measured depth is decreased. If the dilution is high and/or the transit is shallow, this can mimic a planetary transit around the target star.
\end{enumerate}

\section{Computation of the stellar centroid time series data}
\label{s:Computation of the stellar centroid time series data}

The twelve \textit{NGTS} telescopes have a combined field of view of almost $100$~sq.deg. Each CCD is a deep depleted 2k$\times$2k Ikon-L produced by Andor, with pixel size of $13.5~\mu$m ($4.97$~arcsec).
The default radius of the circular photometric aperture is $3$~pixel, covering a total area on sky of $700$~sq.arcsec.
For all observations, the survey employs the {\scshape donuts} auto-guiding algorithm developed by \cite{McCormac2013}, which ensures the telescopes stay centred on target over the course of one night.

The centre of aperture per exposure, $\vec{x}_\mathrm{apt} (t)$, is determined by a global fit to all reference stars in the field of view. 
Thereby, the high-precision auto-guiding minimises random scatter of these aperture positions to ${\sim}0.1$~pixel between subsequent exposures, and a total drift of $< 1$~pixel over multiple hours.
In theory, the centre of flux per exposure, $\vec{x}_\mathrm{flux} (t)$, is equal to $\vec{x}_\mathrm{apt} (t)$ in the case of isolated stars with a Gaussian point spread function and perfect alignment of the aperture mask.
However, in the presence of blended objects or stray light, the two are offset from each other. This offset depends on the magnitude and position of the background object.

Both $\vec{x}_\mathrm{apt} (t)$ and $\vec{x}_\mathrm{flux} (t)$ are computed for each exposure in the \textit{NGTS} pipeline using {\scshape casutools}\footnote{http://casu.ast.cam.ac.uk/surveys-projects/software-release} \citep{Irwin2004}.
We introduce the centroid as a relative value, $\xi$, relating the two:
\begin{equation}
\vec{x}_\mathrm{flux} (t) = \vec{x}_\mathrm{apt} (t) + \vec{\xi} (t).
\end{equation}
As a result, the centroid is automatically corrected for any global drift of the grid of apertures across the CCD, and is representing the remaining (local) residuals.
Note that $\vec{\xi} (t)$ is a time series containing centroid measurements for all exposures.

\subsection{Pre-whitening the flux-centroid time series}
\label{ss:Pre-whitening the centroid time series}

\begin{figure}
 \includegraphics[width=\columnwidth]{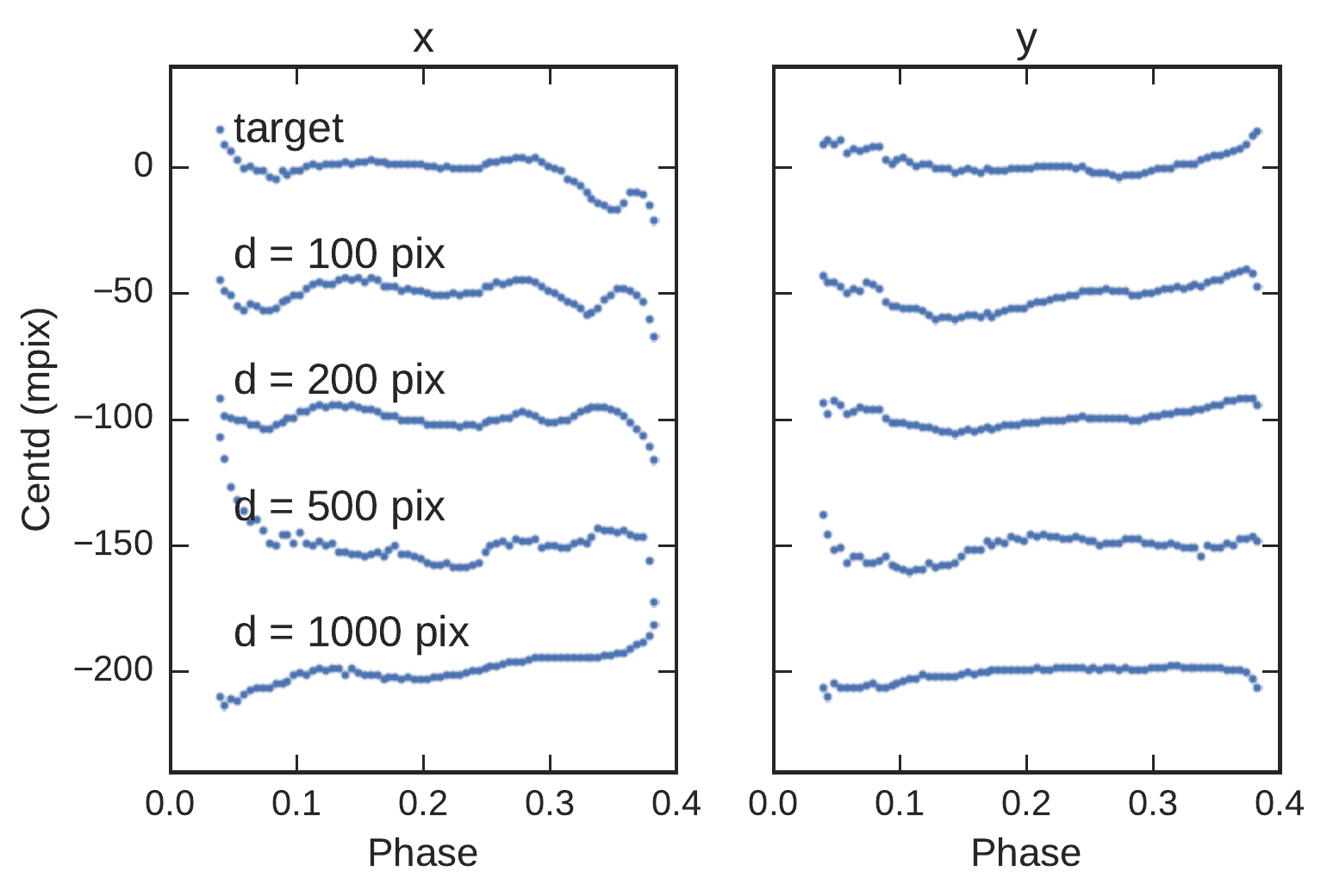}
 \caption{The similarity of the centroid motion between objects decreases with distance on the CCD. Examples show the centroid data of randomly selected objects. This data was collected over the course of four months and phase folded on a sidereal day period.}
 \label{fig:CENTD_neighbours_correlation_phased}
\end{figure}

\textit{NGTS} typically observes each field down to an elevation of $30^{\circ}$ above the horizon. Given the large (${\sim}3^{\circ}$) field-of-view (FOV) there is a $\sim5$~arcsec difference in the atmospheric refraction between the higher and lower elevation sides of the image. This difference ranges from $0.5$~pixels at the zenith, to $1.75$~pixels at an elevation of $30^{\circ}$ (see Eq. G in \citealt{Bennet1982}), with a pixel size of $4.97$~arcsec. This effect is temperature and pressure dependent and acts along the parallactic angle, which rotates as the field crosses the sky. Additionally, small amounts of field rotation may occur due to residual polar misalignment of each \textit{NGTS} telescope. Hence the centroids display a systematic low-amplitude drift over the course of each night (see Fig.~\ref{fig:CENTD_target_nights}). The strength of this effect varies across the image and may differ night-to-night, but is correlated between neighbouring objects. The correlation between objects decreases as the distance between them increases (see Fig.~\ref{fig:CENTD_neighbours_correlation_phased}). The \textit{NGTS} autoguider aims to fix the global average position of the field to the same sub-pixel level, and can hence not correct for this. 
To address this, we instead follow a three-step approach to correct the centroid motion for each target star:
 
 \begin{figure}
 \includegraphics[width=\columnwidth]{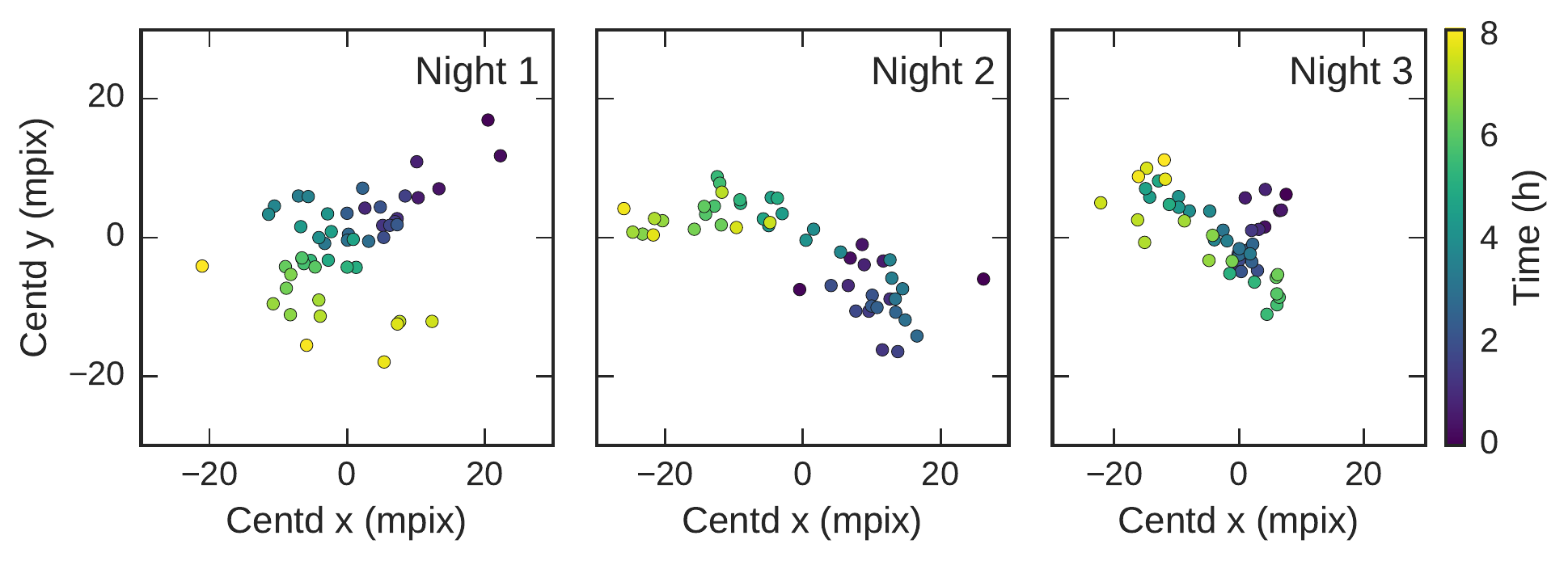}
 \caption{The centroid motion varies between different observing nights. Centroid values are shown in milli-pixel (mpix). The colour coding illustrates the time from the start of the observations on a given night.}
 \label{fig:CENTD_target_nights}
\end{figure}

\begin{enumerate}
\item \textit{Flattening}: we compute the median centroid of each night and use it to correct the night-to-night offset, in order to combine all nights together.
\item \textit{Detrending}: we use the centroid correlation between neighbouring objects to pre-whiten the centroid time series of our planet candidates. 
For a given target, we select $N_\mathrm{ref}$ reference stars. We perform a least square fit to determine which linear combination of these resembles the target's centroid curve best, and remove this trend:
\begin{equation}
\xi_{\mathrm{target, detrended}} = \xi_{\mathrm{target, raw}} - \sum_i^{N_\mathrm{ref}} c_{\mathrm{i, raw}} \cdot\xi_{\mathrm{i, raw}}.
\end{equation}
Here, $\xi_{\mathrm{i, raw}}$ is the centroid data of the $i$-th reference star, and  $c_{\mathrm{i, raw}}$ is the scale parameter that is fitted for.
When selecting reference stars, we only regard objects within a certain distance $d_\mathrm{max}$ (in pixel) from the target on the CCD, based on our observations of decreasing correlation with distance (see Fig. \ref{fig:CENTD_neighbours_correlation_phased}). 
Further, we pre-select the most correlated objects to decrease the number of free fit parameters. 
We phase-fold the centroid time series on the transit period, exclude the in-transit data, and calculate Pearson's correlation coefficient of the target with each selected neighbour. The $N_\mathrm{ref}$ most correlated objects are selected.
Finally, to choose reference stars which are less affected by residuals of the sky background ($\lesssim 50$~ADU/s) subtraction and to avoid saturated stars ($\gtrsim 50000$~ADU/s), only objects with flux of ${500-10000}$~ADU/s are included.

\item \textit{Sidereal day correction}: 
The observing pattern of ground-based surveys can lead to systematic noise on the period of a sidereal day. As our centroid detrending is applied to data that has been phase-folded on the transit period, residuals of sidereal day systematics may be present in long-period systems. To further enhance our algorithm, we phase-fold the centroid time series of the target on the mean period of a sidereal day and perform a moving average fit to correct for any remaining trends. We only consider data outside of the primary or secondary eclipses. Hence, the correction does not affect the transit signal. Generally, the effect of the sidereal day correction is marginal, as most \textit{NGTS} targets are found at short periods.
\end{enumerate}

\subsection{Effect of the different detrending steps}
\label{ss:Effect of the different detrending steps}

\begin{figure}
 \includegraphics[width=\columnwidth]{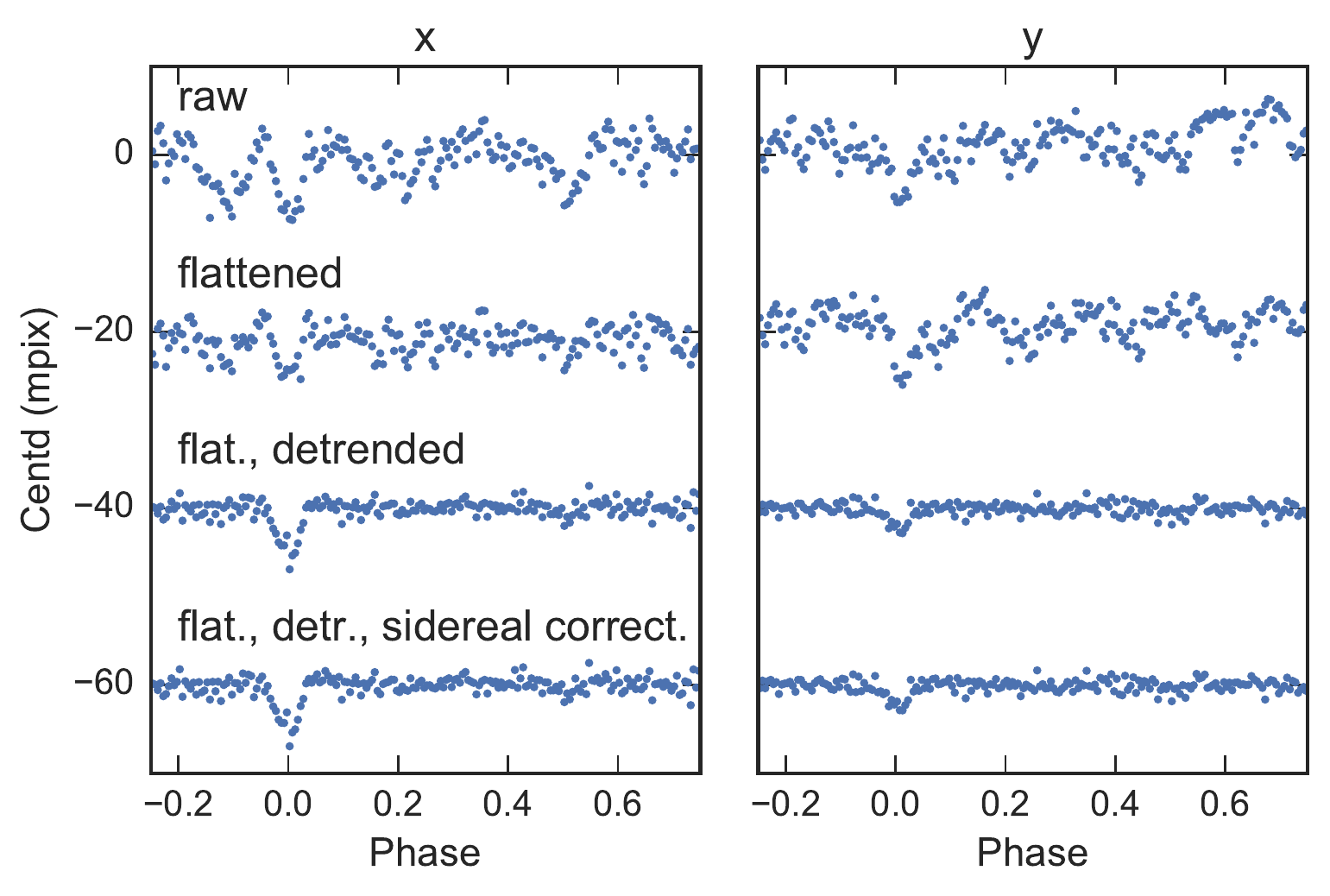}
 \caption{
 Improvement of the centroiding systematics with each detrending step (see section~\ref{ss:Pre-whitening the centroid time series}) on the example of NG0409-1941 020057. Shown are the centroid time series phase-folded on the period of the transit-like signal at $1.61$ days. 
 First, we correct the night-to-night offsets in the time series by subtracting the median value per night (`flattened'). Second, the centroid data is detrended by a reference signal calculated from their most correlated neighbours. Last, a sidereal day correction is applied.
 Centroid data in x and y direction are shown in the left and right column of the figure, respectively.
 The time series are offset from each other by $20$~milli-pixel on the vertical axis for clarity.}
 \label{fig:CENTD_detrending_viridis}
\end{figure}

Fig.~\ref{fig:CENTD_detrending_viridis} demonstrates the effect of our centroid detrending steps for a chosen \textit{NGTS} planet candidate (see section~\ref{ss:Pre-whitening the centroid time series}). 
First, the night-to-night offsets in the time series are removed (`flattening'). Second, we detrend the centroid phase-folded on the transit period, using
a reference signal computed from the most correlated neighbours.
This leads to a clear improvement, removing almost all systematics. 
Finally, a sidereal day correction is applied.
Since NG0409-1941 020057 has a period of $1.61$~days (see section~\ref{sss:NG 0409-1941 020057}), which is close to the sidereal day period, these systematics have mostly been removed before this stage, such that their effects are negligible. 
Note that long-period transits, however, will profit from this additional step. 
After detrending, a clear centroid signal remains at phase $0$, indicating the presence of a background object in the aperture (see section \ref{s:Introduction}).

\subsection{Global performance: sub-milli-pixel precision}
\label{ss:Achieved centroid precision}

To assess the global performance of our technique, we mimic the centroiding process for planet candidates on $\sim$1200 targets from a typical \textit{NGTS} field. For each star, we randomly uniformly draw a period between $0.8$ and $15$ days, on which we phase-fold the centroid time series.
This range is based on the minimum value used for the Kepler planet occurrence rates in \cite{Fressin2013} and the period sensitivity of \textit{NGTS}.
We select stars with flux counts of ${500-10000}$~ADU/s, to avoid influence of the sky background (${\sim}50$~ADU/s) and saturation ($>50000$~ADU/s).

\begin{figure}
 \includegraphics[width=\columnwidth]{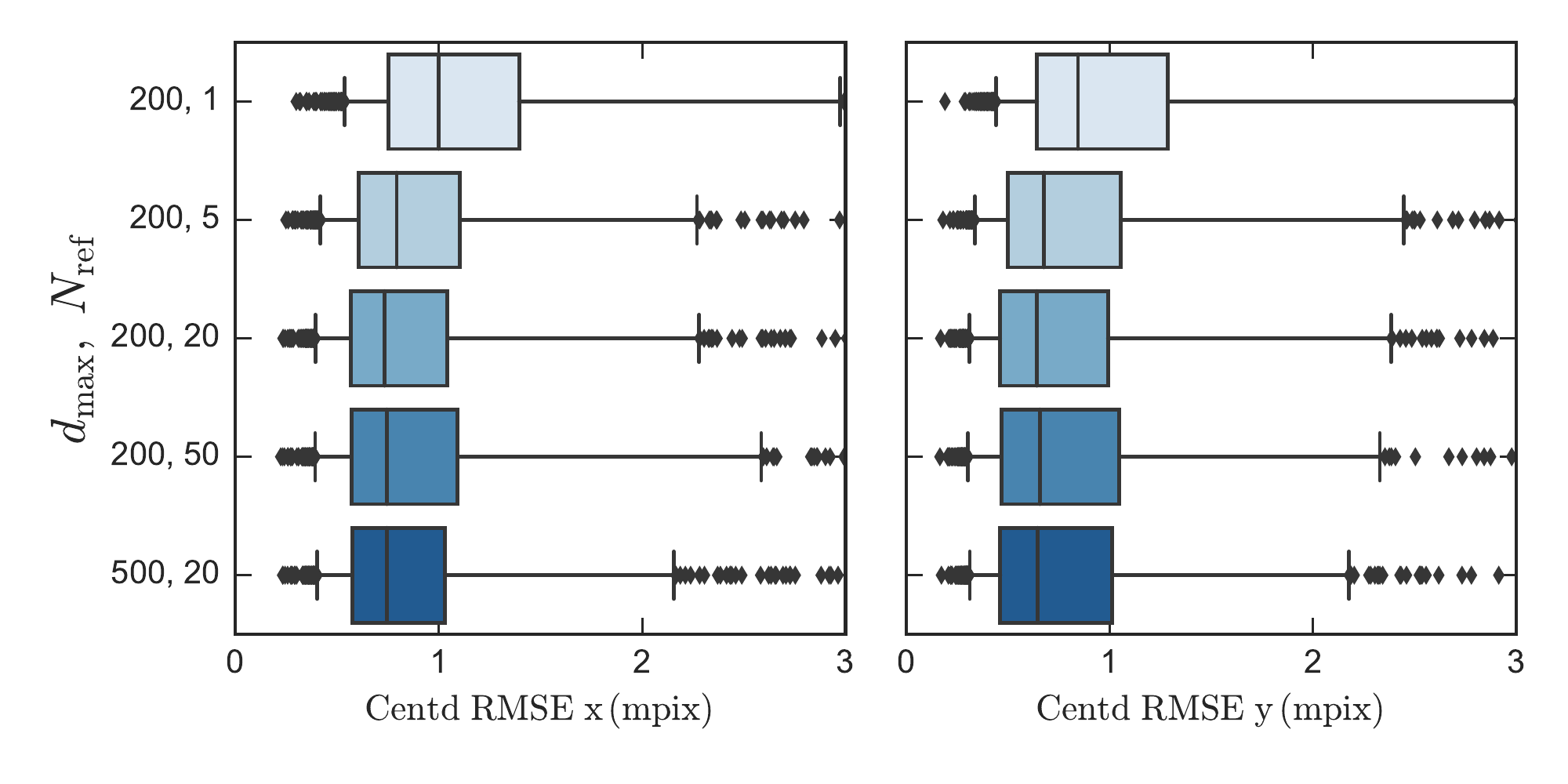}
 \caption{Comparison of different centroid detrending settings. Shown is the root mean squared error (RMSE) of the phase-folded data after detrending. Boxes represent the median, and 25th and 75th percentile of all objects. Whiskers display the 5th and 95th percentile, and outlying objects are plotted as symbols. A search radius of $d_\mathrm{max} \approx 200$ pixel and choice of $N_\mathrm{ref} \approx 20$ reference objects is the best compromise between a high precision and computational efficiency.}
 \label{fig:CENTD_Detrending_methods_evaluation}
\end{figure}

We test the impact of different settings for the maximum distance on the CCD, $d_\mathrm{max}$, and maximum number of reference stars, $N_\mathrm{ref}$ (Fig.~\ref{fig:CENTD_Detrending_methods_evaluation}). 
Selecting the most correlated neighbour as the sole reference object already leads to an average milli-pixel precision. 
Including more reference objects further increases this precision, yet saturates at $N_\mathrm{ref} \approx 20$.
Widening the search radius does not yield any improvement, as the correlation of the centroid systematics decreases with distance (see section~\ref{ss:Pre-whitening the centroid time series} and Fig.~\ref{fig:CENTD_neighbours_correlation_phased}).
We find an optimal performance for $d_\mathrm{max} \approx 200$ pixel and $N_\mathrm{ref} \approx 20$. 
In theory, additional reference objects add additional information, but can also lead to over-fitting or converging to a local minimum. 
In any case, pre-selecting a limited number of the most correlated reference stars is advantageous for computational efficiency.

\begin{figure}
 \includegraphics[width=\columnwidth]{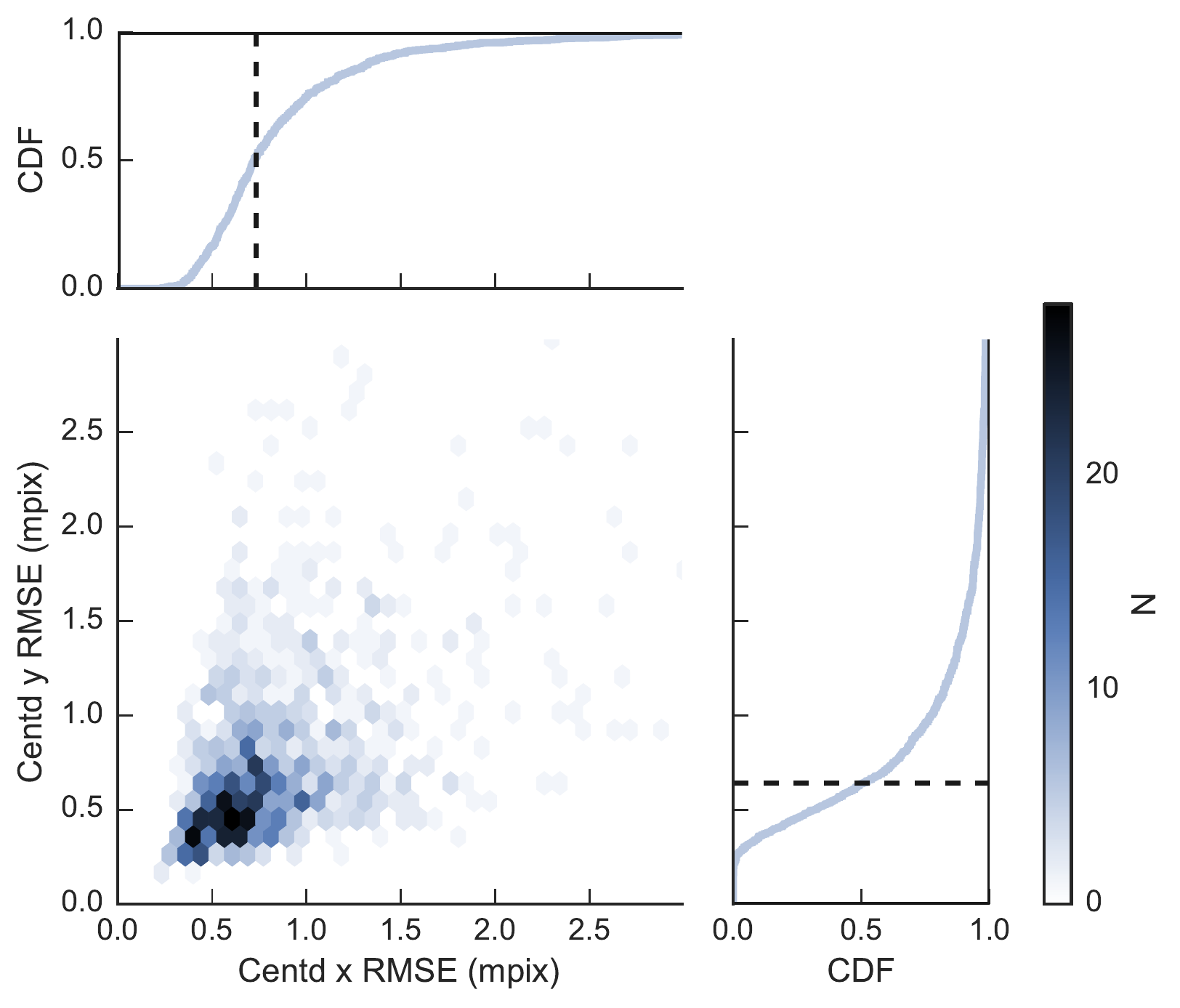}
 \caption{Achieved centroid precision for an analysis run over ${\sim}1200$ stars from a typical \textit{NGTS} field. Each star is given a period, which is randomly uniformly drawn between $0.8$ and $15$ days, and the centroid time series is phase-folded on the respective period. Shown is the root mean squared error (RMSE) of the phase-folded data in x and y direction after detrending, as well as the cumulative distribution function (CDF) in each direction.}
 \label{fig:Precision_hist}
\end{figure}

Fig.~\ref{fig:Precision_hist} illustrates the remaining root mean squared error (RMSE) of the phase-folded centroid data after detrending. 
We achieve an RMSE of $<1$~milli-pixel for $73$ and $75$~per-cent of all targets in the x and y direction, respectively. $61$~per-cent show a precision of $<1$~milli-pixel in both directions at the same time. 
We achieve an average centroid precision of $\sim0.75$~milli-pixel in both directions, and as low as $0.25$~milli-pixel for individual targets.

\begin{figure}
 \subfigure{\includegraphics[width=\columnwidth]{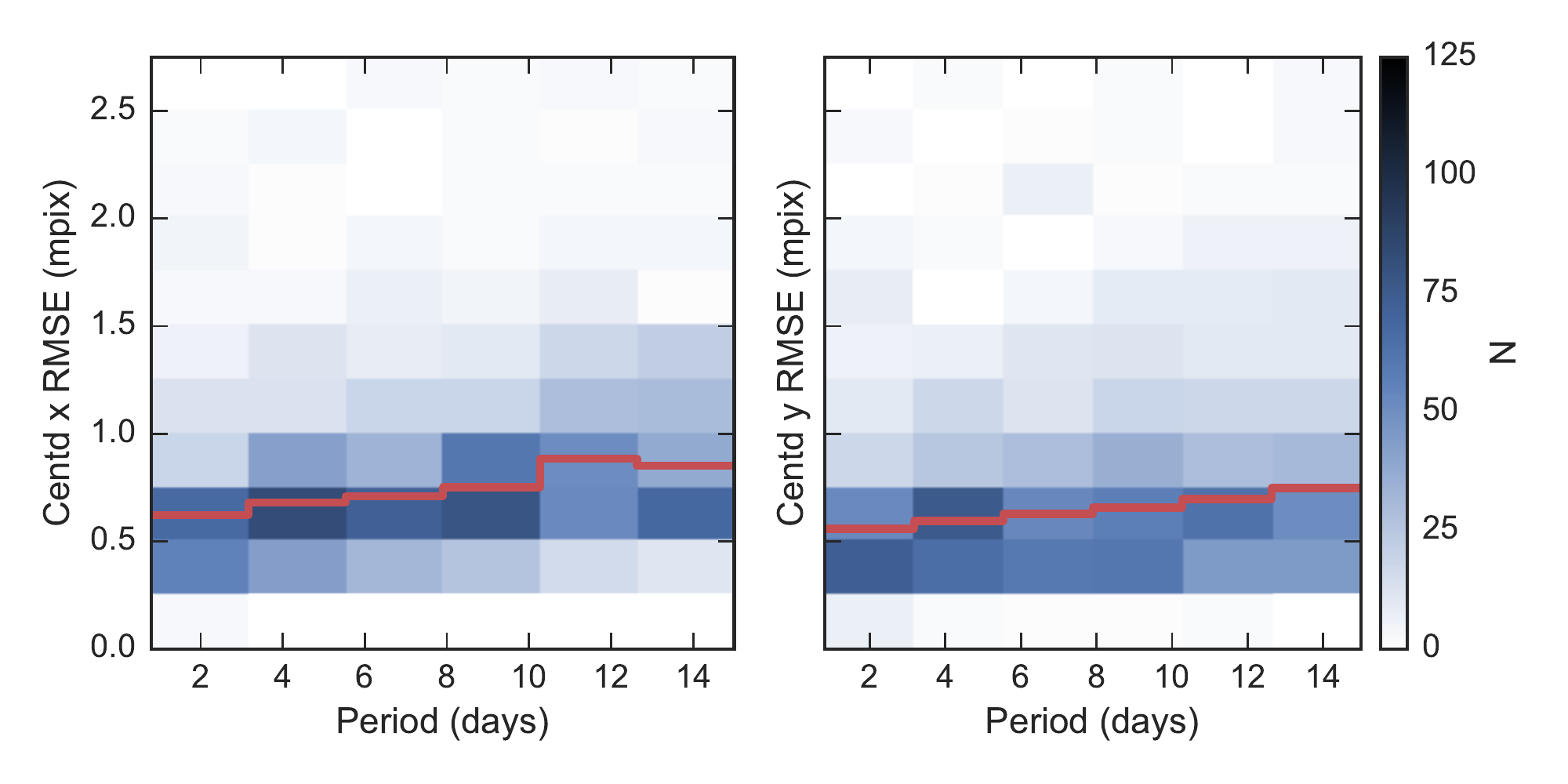}}
 \subfigure{\includegraphics[width=\columnwidth]{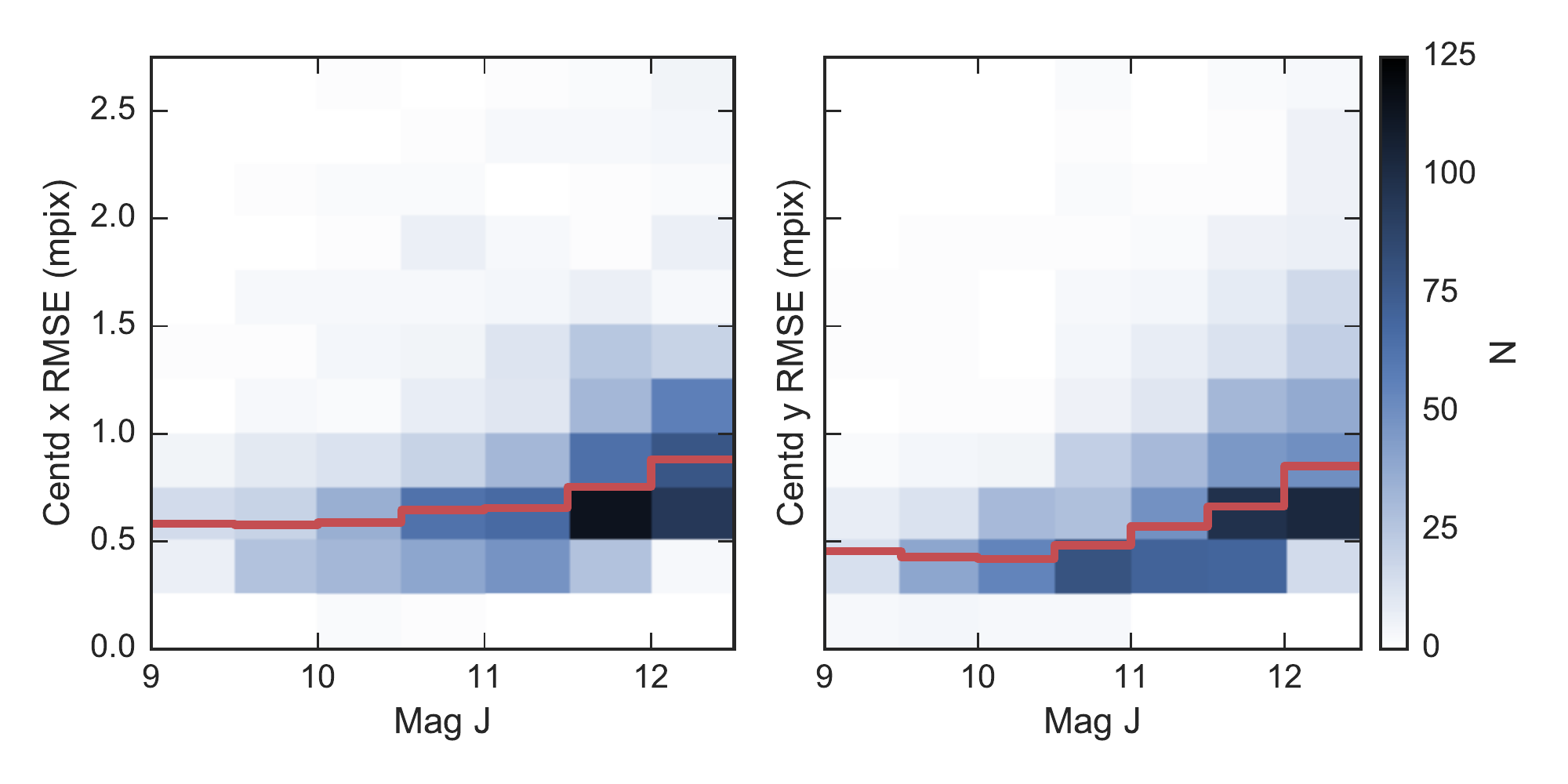}}
 \caption{Dependency of the centroid precision versus period of the transit-like signal (upper panel) and the J-magnitude of the host star taken from \textit{2MASS} (lower panel), shown for ${\sim}1200$ targets from a typical \textit{NGTS} field. Solid red lines indicate the median centroid RMSE for each period and magnitude bin. Long period signals show slightly higher noise in the phase-folded centroid curves. Fainter stars are more influenced by the sky background, leading to increased noise in the phase-folded centroid curves.}
 \label{fig:CENTD_vs_X_hist2D}
\end{figure}

We identify an increase of the achieved centroid precision for shorter periods (Fig.~\ref{fig:CENTD_vs_X_hist2D}, upper panel). This is a direct consequence of the white noise statistics and the amount of data that can be binned up in the phase-folded centroid curve. 
Similarly, the centroids of fainter host stars are stronger influenced by sky background variations, leading to higher noise in the phase-folded centroid curves (Fig.~\ref{fig:CENTD_vs_X_hist2D}, lower panel).

\section{Identification of centroid shifts caused by blended sources}
\label{ss:Identifying centroid shifts caused by blended sources}

To identify centroid shifts caused by blended sources in the aperture of a planet candidate, we phase-fold the detrended centroid time series on the period of the respective transit feature. We compare the flux and centroid phase curves using four methods:

\begin{enumerate}
\item Manual inspection of the phase-folded flux and centroid curves (see panel A in Fig.~\ref{fig:CENTD_identification_NG0522-2518_017220_TEST18} and \ref{fig:CENTD_identification_NG0409-1941_020057_TEST18}).
This allows a qualitative investigation of the noise in each time series, and the identification of any systematic features that might mimic or hide a correlation between flux and centroid.

\item Pearson's correlation for a rolling window\footnote{known as rolling, windowed, or sliding-window correlation} (see panel B in Fig.~\ref{fig:CENTD_identification_NG0522-2518_017220_TEST18} and \ref{fig:CENTD_identification_NG0409-1941_020057_TEST18}).
The window size should be longer than the signal width. In praxis, we employ multiple window sizes and include all results into our further analyses. For the purpose of this paper, we demonstrate our analyses using a window size of $0.25$ in phase.
To start, we place this window centred on the transit and calculate Pearson's correlation coefficient between the flux and centroid data. We then slide the window across the data, repeating the measurement for each position.

\item Cross-correlation (see panel C in Fig.~\ref{fig:CENTD_identification_NG0522-2518_017220_TEST18} and \ref{fig:CENTD_identification_NG0409-1941_020057_TEST18}).
We calculate Pearson's correlation coefficient between the phase-folded flux and each centroid curve. We then shift one series of data against the other (with periodic bounds), repeating the measurement for each position. This is equivalent to the cross-correlation function widely used in astronomy, but normalised to a range of $-1$ to $+1$.

\item Hypothesis tests and manual inspection of the rain plots (see panel D in Fig.~\ref{fig:CENTD_identification_NG0522-2518_017220_TEST18} and \ref{fig:CENTD_identification_NG0409-1941_020057_TEST18}).
Rain plots were successfully used to qualitatively identify correlations between the flux and centroid time series for \textit{Kepler} (see e.g. \citealt{Batalha2010}). While a single target undergoing an eclipse would `rain' straight down (local centre of flux is unaffected), a blended object causes a trend (`wind') sideways. 
In our analyses for \textit{NGTS}, we extract the in-transit data as a subset. We set the Null Hypothesis that this in-transit centroid data is distributed around the mean of the out-of-transit centroid data, i.e. around $0$.
We perform statistical hypotheses tests, a two-tailed T-test and two-tailed binomial test, and record their p-value. For a chosen significance level, e.g. $alpha = 0.01$, we test if the the Null Hypothesis can be rejected.
The two-tailed tests are employed to verify both if the mean is significantly greater or significantly smaller than $0$ (whereas a one-tailed test would only test for one direction). 

\end{enumerate}

\subsection{Analysis of blended systems}
\label{ss:Analysis of blended systems}

A detailed study of the centroid signal is needed to determine which scenario causes the transit-like feature (dilP, dilEB, BP, or BEB; see section~\ref{s:Introduction}).
At this stage in the vetting process we assume that the period has been identified correctly.

In the first step of the centroid analysis, any information on visually identified nearby objects must be used to identify blended sources. 
We therefore inspect the \textit{NGTS} images and cross-match with existing catalogues including \textit{Gaia} DR1 \citep{Gaia,GaiaDR1} and 2MASS \citep{2MASS}.
If the local centre of flux shifts away from (towards) the centre of the target star, this indicates that the target star (background object) is decreasing in brightness, i.e. undergoes the eclipse (see Fig~\ref{fig:sketch}).
However, as discussed in section~\ref{s:Introduction}, this alone does not disqualify a planet scenario. Only a detailed model of the flux and centroid time series simultaneously, taking the dilution factor into account, can guide in this question by establishing the likelihood of all astrophysical parameters.

In the following, we model the aperture to contain two sources of light, one that is constant ($\mathrm{c}$) and one that is eclipsing ($\mathrm{e}$). This implicitly models multiple background objects as originating from a single source, located at their centre of light in the aperture.
Consequently, $F_{\mathrm{e}}(t)$ is the flux time series of the eclipsing object alone, while $F_{\mathrm{c}}(t) = F_{\mathrm{c}}$ denotes the constant object.
The time series of the total flux in the aperture is
\begin{equation}
F_\mathrm{sys}(t) = F_\mathrm{e}(t) + F_\mathrm{c}.
\end{equation}

\noindent We define the dilution factor out of transit as
\begin{equation}
\label{eq:DIL}
D_0 = 1 - \frac{F_{\mathrm{e}}(t_0)}{F_{\mathrm{sys}}(t_0)},
\end{equation}
where $t_0$ is a chosen time out-of-transit.

We have to distinguish between two cases: an eclipsing background object (BP or BEB; section~\ref{sss:Constant target, eclipsing background object}) and a constant background object (dilP or dilEB; section~\ref{sss:Eclipsing target, constant background object}). We consider these two cases in turn in the following two subsections.

\subsubsection{Constant target, eclipsing background object}
\label{sss:Constant target, eclipsing background object}

The target is constant (located at $\vec{x}_\mathrm{c}$) and the signal comes from an eclipsing background source that is offset (located at $\vec{x}_\mathrm{e}$). 
We set the origin of the coordinate system to the target's position, hence $\vec{x}_\mathrm{c} = 0$.
The centroid out of transit, $\vec{\xi}_0$, is then given following Eq.~\ref{eq:xi} and \ref{eq:DIL} as:
\begin{align}
\vec{\xi}_0 &= \frac{F_\mathrm{c}(t_0) \vec{x}_\mathrm{c} + F_\mathrm{e}(t_0) \vec{x}_\mathrm{e}}{F_\mathrm{sys}(t_0)}
 = \frac{F_\mathrm{e}(t_0) \vec{x}_\mathrm{e}}{F_\mathrm{sys}(t_0)} \nonumber \\
 &= \left( 1 - D_0 \right) \vec{x}_\mathrm{e}
\end{align}
The centroid at any time is then given as
\begin{align}
\vec{\xi}(t) &= \frac{F_\mathrm{c}(t) \vec{x}_\mathrm{c} + F_\mathrm{e}(t) \vec{x}_\mathrm{e}}{F_\mathrm{sys}(t)} - \vec{\xi}_0 \nonumber \\
&= \left( \frac{F_\mathrm{e}(t)}{F_\mathrm{sys}(t)} - 1 + D_0 \right) \vec{x}_\mathrm{e}
\end{align}
For normalised lightcurves, we can express this as
\begin{align}
F_\mathrm{e}(t) &= F_\mathrm{e}^\mathrm{norm}(t) \cdot F_\mathrm{e}(0) \\
F_\mathrm{sys}(t) &= F_\mathrm{sys}^\mathrm{norm}(t) \cdot F_\mathrm{sys}(0) \\
\dfrac{F_\mathrm{e}(t)}{F_\mathrm{sys}(t)} 
&= \dfrac{F_\mathrm{e}^\mathrm{norm}(t)}{F_\mathrm{sys}^\mathrm{norm}(t)} \cdot \left( 1 - D_0 \right) \\
\vec{\xi}(t) &= \left( \frac{F_\mathrm{e}^\mathrm{norm}(t)}{F_\mathrm{sys}^\mathrm{norm}(t)} - 1 \right) \cdot \left(1 - D_0 \right) \vec{x}_\mathrm{e}.
\label{eq:BEB}
\end{align}

\subsubsection{Eclipsing target, constant background object}
\label{sss:Eclipsing target, constant background object}

The target undergoes the eclipse (located at $\vec{x}_\mathrm{e}$) and is diluted by a constant background source that is offset (located at $\vec{x}_\mathrm{c}$). 
We set the origin of the coordinate system to the target's position, hence $\vec{x}_\mathrm{e} = 0$.
The centroid out of transit, $\vec{\xi}_0$, is then given following Eq.~\ref{eq:xi} and \ref{eq:DIL} as:
\begin{align}
\vec{\xi}_0 &= \frac{F_\mathrm{c}(0) \vec{x}_\mathrm{c} + F_\mathrm{e}(0) \vec{x}_\mathrm{e}}{F_\mathrm{sys}(0)}
 = \frac{F_\mathrm{c}(0) \vec{x}_\mathrm{c}}{F_\mathrm{sys}(0)} \nonumber \\
 &= D_0 \vec{x}_\mathrm{c}
\end{align}
The centroid at any time is then given as
\begin{align}
\vec{\xi}(t) &= \frac{F_\mathrm{c}(t) \vec{x}_\mathrm{c} + F_\mathrm{e}(t) \vec{x}_\mathrm{e}}{F_\mathrm{sys}(t)} - \vec{\xi}_0 \nonumber \\
&= \left( \frac{F_\mathrm{c}(t)}{F_\mathrm{sys}(t)} - D_0 \right) \vec{x}_\mathrm{c}
\end{align}
For normalised lightcurves, we can express this as
\begin{align}
F_\mathrm{c}(t) &= F_\mathrm{c}^\mathrm{norm}(t) \cdot F_\mathrm{c}(0) =  1 \cdot F_\mathrm{c}(0)\\
F_\mathrm{sys}(t) &= F_\mathrm{sys}^\mathrm{norm}(t) \cdot F_\mathrm{sys}(0) \\
\dfrac{F_\mathrm{c}(t)}{F_\mathrm{sys}(t)} 
&= \dfrac{1}{F_\mathrm{sys}^\mathrm{norm}(t)} \cdot D_0 \\
\vec{\xi}(t) &= \left( \frac{1}{F_\mathrm{sys}^\mathrm{norm}(t)} - 1 \right) \cdot D_0 \vec{x}_\mathrm{c}
\label{eq:BCS}
\end{align}

\subsubsection{Dependency of the blend position on transit parameters}
\label{sss:Blend position in dependency of transit parameters}

Rearranging the previous equations allows us to express the position of the blended background object in terms of the actually measured transit depth of the system, $\delta_{sys}$:
\begin{equation}
\label{eq:loc_blend}
\mathrm{{\vec{x}}_{blend}} = \pm ~\mathrm{\vec{\xi}_{max}} \cdot \left( \frac {\delta_{sys}} {1 - \delta_{sys}} \cdot D_0 \right)^{-1}. 
\end{equation}
The sign depends on whether the blended background object is the constant (+) or variable (-) source in the aperture.

\subsubsection{Bayesian analysis}
\label{sss:Bayesian analysis}

A Bayesian analysis enables us to explore complex parameter spaces and robustly estimate posterior likelihoods for each parameter. Specifically, any information on visually identified objects can be used as priors, such as the position of the blended background object. 

We base our fitting model on the {\scshape eb}\footnote{\url{https://github.com/mdwarfgeek/eb}, online 05 Feb 2017}\footnote{see Acknowledgements} module by \citet{Irwin2011}, which incorporates the possibility of a dilution term.
We establish a maximum likelihood function based on a simultaneous fit of the phase-folded flux and centroid time series following Eq.~\ref{eq:BEB} and \ref{eq:BCS} (depending on which model we chose to fit). 
The free parameters in our model are the relative CCD position of the blended background object ($\mathrm{{\vec{x}}_{blend}}$), the dilution factor $D_0$, as well as the standard parameters of an EB model, namely the surface brightness ratio $J$, ratio of the sum of the radii over the orbital distance $(R_1 + R_2)/a$, the radius ratio $R_2/R_1$, and cosine of the inclination $\cos{i}$.
We also include offset terms for the normalised flux and centroid time series, denoted $F_0$, $\xi_{x,0}$, and  $\xi_{y,0}$. These correct for any remaining offset after the normalisation, and are usually negligible.
Additionally, the error bars on the flux and centroid data are free parameters and are determined by the fit, $\sigma(F)$, $\sigma(\xi_x)$, and $\sigma(\xi_y)$.
Where there is no prior information, we choose uniform priors.
For systems which pass through the vetting to this stage, we expect low-brightness ratio systems that do not show significant secondary eclipses, hence uniform priors are generally expected to be uninformative. 

We fix the period and epoch to the values determined by the \textit{NGTS} candidate pipeline. The other parameters of the {\scshape eb} model (limb darkening, gravity darkening, and reflectivity) are left at their standard values. Both stars are assumed to be without spots.
The {\scshape eb} model can resemble the \citet{Mandel2002} planet transit model if surface brightness ratio, mass ratio, light travel time and reflectivity are zero. Hence, we can readily model all scenarios (dilP, dilEB, BP, BEB; see section~\ref{s:Introduction}).

To find the best fit, we follow a two-step approach. First, we employ a differential evolution algorithm \citep{Storn1997} to explore the parameter space and find a global optimisation. This uses an iterative approach in which different populations of solutions are compared and the best fit is kept. As it does not rely on gradient methods, it is robust against local minima.
We implement the {\scshape scipy} \citep{Jones2001} distribution of the algorithm.

Second, we use the result of the differential evolution as the initial guess for an Markov chain Monte Carlo (MCMC) algorithm and adopt priors, to re-fine the fit and establish the likelihood of our parameters.
We implement the {\scshape emcee} \citep{Foreman-Mackey2013} package.
Initially, the walkers are distributed following a Gaussian distribution around the initial guess with the standard deviation being $1\%$ of the given data range. We first perform several MCMC runs with 10000 steps each, and test for convergence using the Gelman-Rubin statistic \citep{Gelman1992}. In the final run, we compute 5000 burn-in steps and 45000 evaluation steps, from which we sample every $n$-th step, whereby $n$ is determined by the maximum of the autocorrelation time of all parameters. The Gelman-Rubin statistics for all parameters lie well below the recommended value $\hat{R}<1.1$, suggesting convergence of the MCMC chains.


\subsection{Case studies}
\label{ss:Case studies}

In the following we employ our centroiding technique on the example of two case studies representing different scenarios. First, we consider NG 0522-2518 017220, which is an eclipsing binary slightly diluted by a visually resolved blend. Second, we apply our analyses to NG 0409-1941 020057, which has no prior visual information, yet can be identified as a strongly diluted background eclipsing binary from a joint MCMC fit of photometric flux and centroid data.

\subsubsection{NG 0522-2518 017220}
\label{sss:NG 0522-2518 017220}

The target NG 0522-2518 017220 was first detected with a period of ${\sim}0.83$ days. Its sinusoidal out-of-eclipse (OOE) variation unveiled that the true signal originates from a primary and secondary eclipse with a period of ${\sim}1.67$ days, which have comparable depths of ${\sim}3$~per~cent and widths of ${\sim}2.6$ hours. With $G=13.6$ in \textit{Gaia} DR1 ($J=12.6$ and $K=12.2$ in 2MASS) the object is bright and well-suited for follow-up and potential characterisation. It is located at \mbox{RA = 05h 23m 31.6s}, \mbox{DEC = -25d 08m 48.4s} and has been identified as \mbox{2MASS 05233161-2508484}, and \mbox{GAIA 2957881682551005056}.

\begin{figure}
 \includegraphics[width=\columnwidth]{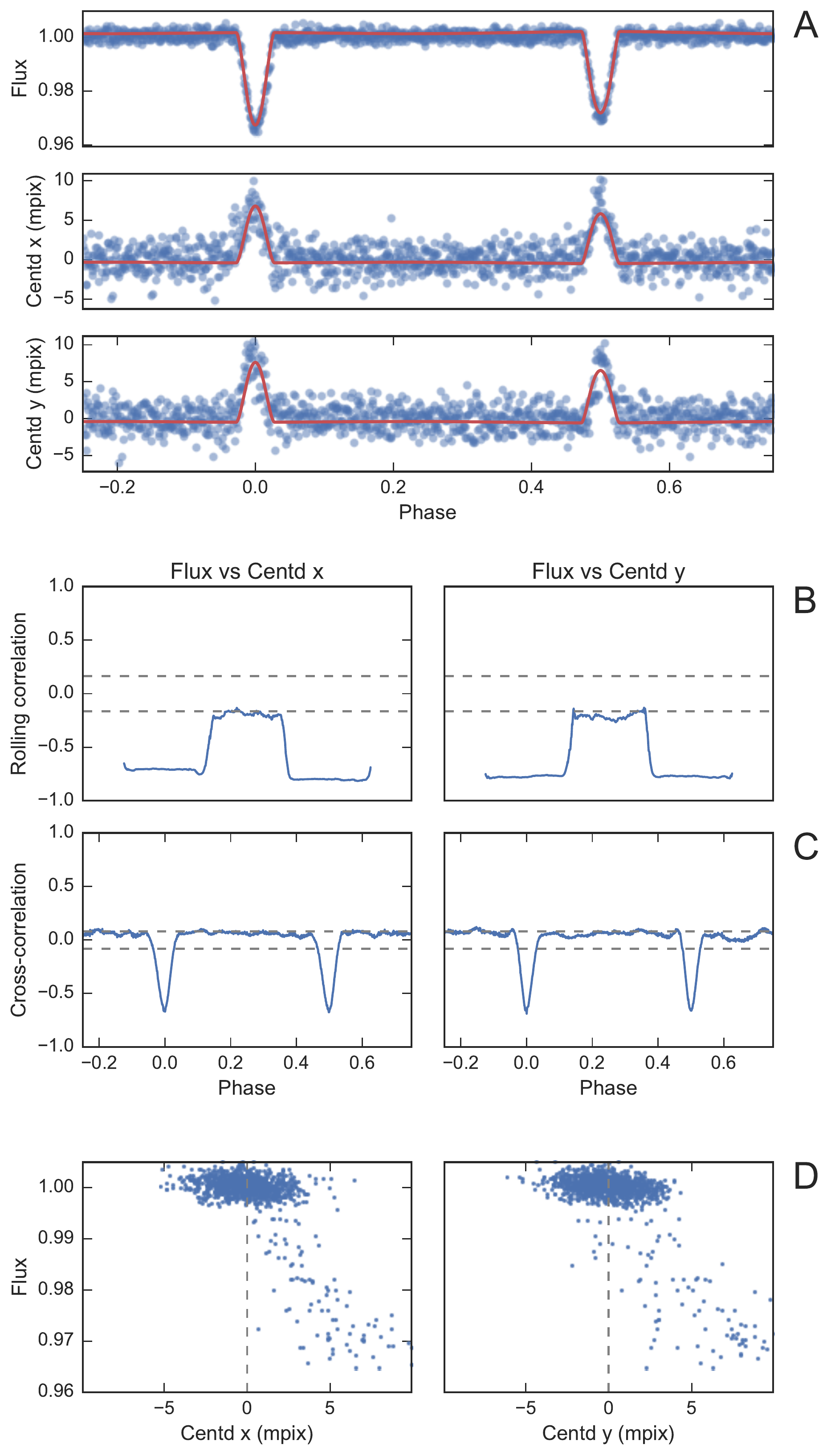}
 \caption{Identification and model fit of a centroid shift correlated to the transit-like signal in NG 0522-2518 017220. A) The correlation is clearly visible in the manual inspection of the flux and centroid phase curves, even though the centroid shift is at the ${\sim}6$~mpix level. Red curves represent the result of the MCMC analysis. B) The rolling correlation analysis shows a significant correlation around phase 0. Dashed lines indicate confidence intervals of $99$~per-cent, calculated as $2.58/\sqrt{w / \Delta t}$ (see \citet{Fisher1921}), whereby $w$ is the window size and $\Delta t$ is the difference in phase between binned points. C) The cross-correlation shows a significant correlation at lag 0 in phase. Dashed lines show confidence intervals of $99$~per-cent calculated as $2.58/\sqrt{N}$ (see \citet{Fisher1921}), whereby $N$ is the total number of binned points. D) Rain plots show a clear trend between flux and centroid shift.}
 \label{fig:CENTD_identification_NG0522-2518_017220_TEST18}
\end{figure}

\begin{figure}
 \includegraphics[width=\columnwidth]{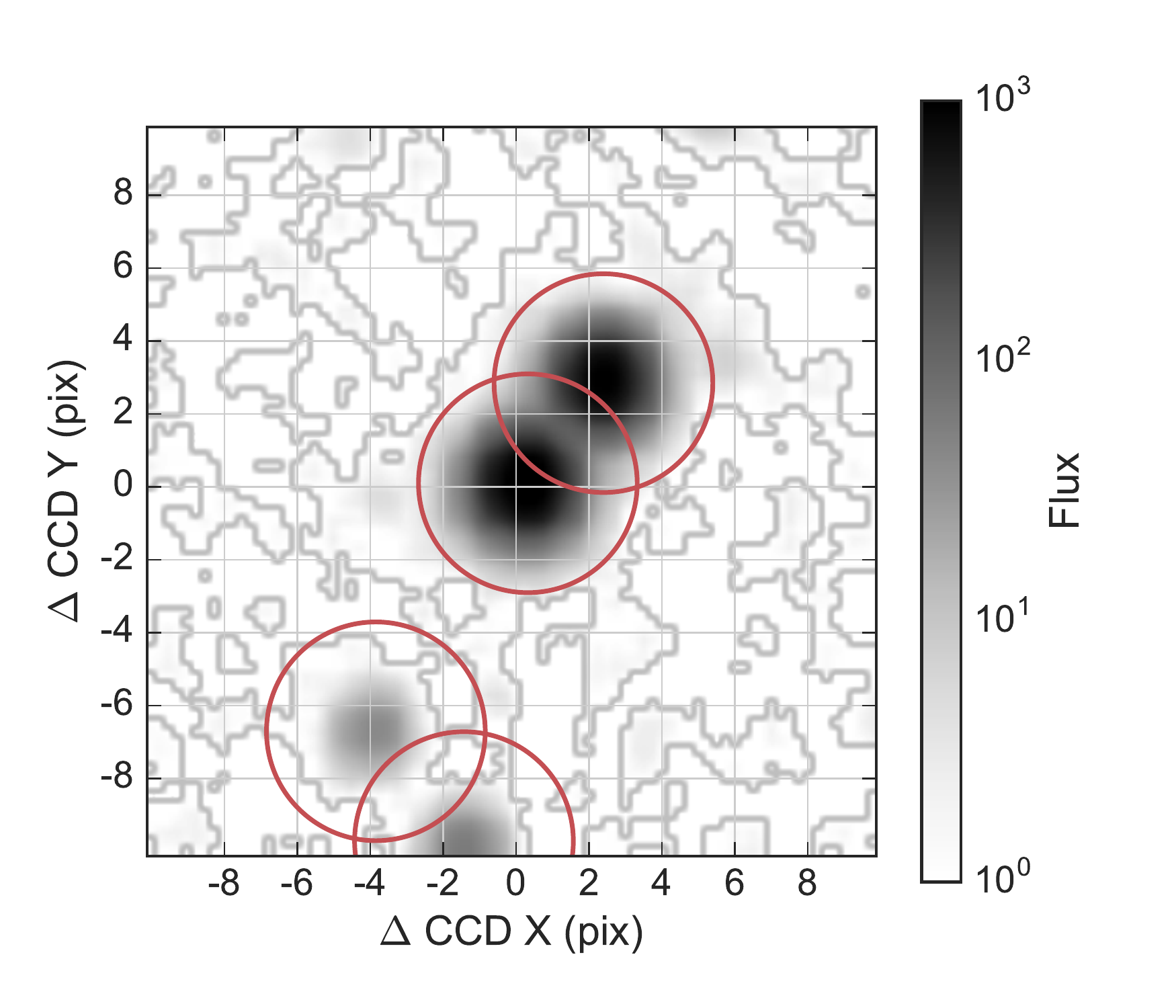}
 \caption{Visual inspection of the \textit{NGTS} sky images for NG 0522-2518 017220 uncovers that a neighbouring object is blending into the photometric aperture of the target (centred in the image). The size of one \textit{NGTS} pixel measures $4.97$~arcsec. Red circles illustrate the photometric aperture radius of 3 pixels.}
 \label{fig:NG0522-2518_017220_TEST16A_astronomy_fit}
\end{figure}

\begin{figure}
 \includegraphics[width=\columnwidth]{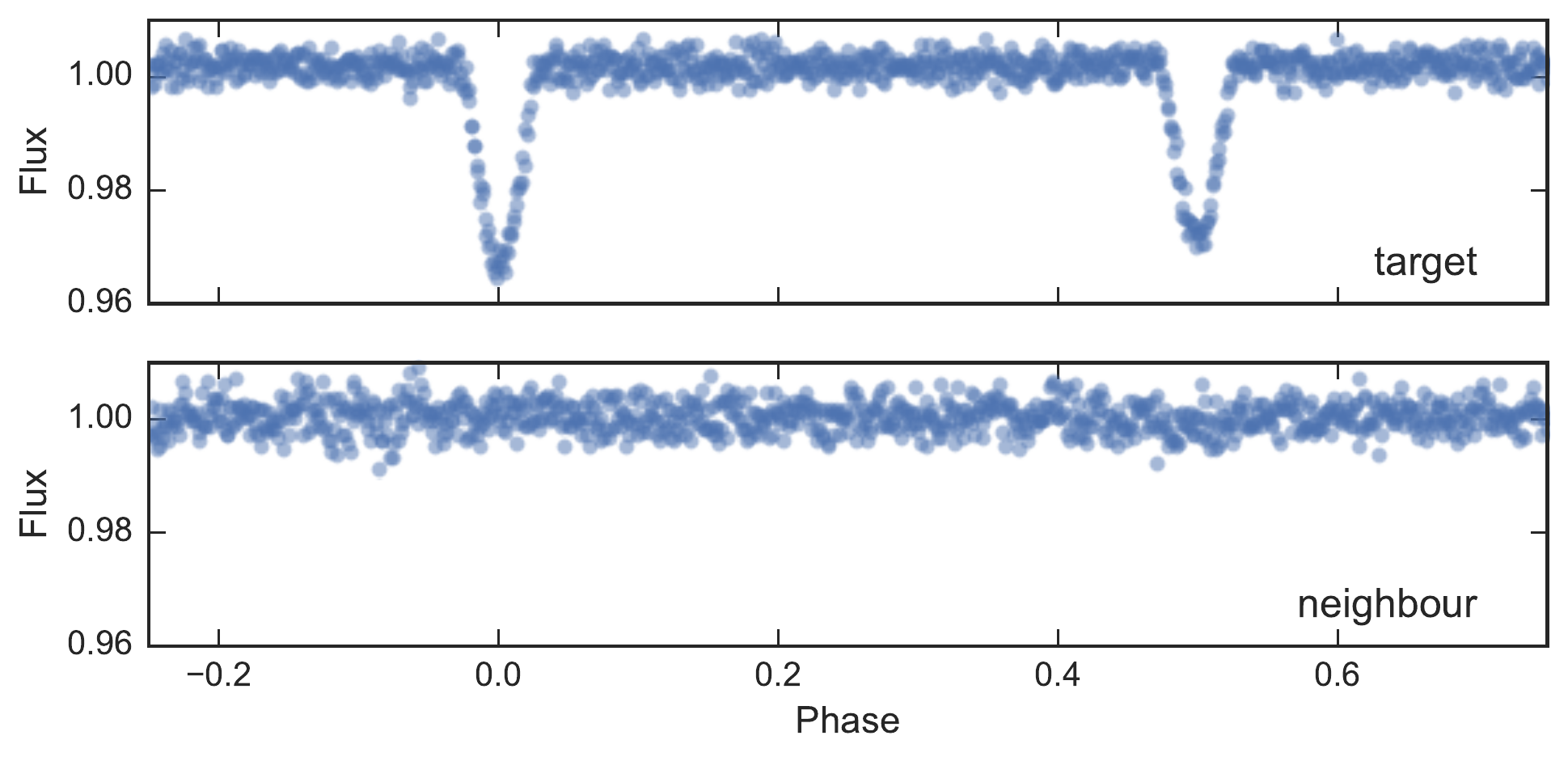}
 \caption{Comparison of the phase-folded lightcurves for NG 0522-2518 017220 and it's blending neighbour. While the primary and secondary eclipse signals are clearly visible in the target, they can not be identified in the blending neighbour. This verifies the outcome of the centroid analysis: the target is undergoing the eclipse, while the neighbouring object is constant.}
 \label{fig:NG0522-2518_017220_vs_017221}
\end{figure}

However, the centroid time series shows clear shifts of $5-8$~mpix in x and y direction, respectively, for both the primary and secondary eclipse (see Fig.~\ref{fig:CENTD_identification_NG0522-2518_017220_TEST18}). All correlation and hypothesis tests confirm a statistically significant centroid shift correlated to the transit signal (see Tab.~\ref{tab:NG 0522-2518 017220}).
On visual inspection, we identify a neighbouring object at ${\sim}18$ arcsec separation, which has a similar brightness at $G=14.0$ ($J=12.7$, $K=12.2$) and partly blends into the target's photometric aperture (see Fig.~\ref{fig:NG0522-2518_017220_TEST16A_astronomy_fit}). 
The centroid shifts into the positive x and y direction, which combined with the visual information suggests that the target is undergoing the eclipse, while the flux from the blended background object is constant (see Fig.~\ref{fig:sketch}). 
A direct comparison of the detrended and phase-folded lightcurves shows that the eclipses are only visible for NG 0522-2518 017220. In the lightcurve of the neighbouring object the signals are diluted beyond the noise level, and hence not detectable (Fig.~\ref{fig:NG0522-2518_017220_vs_017221}). This verifies the conclusions drawn from the centroid analysis.

Using the \textit{GAIA} DR1 magnitude and models for the \textit{NGTS} point-spread-function and bandpass, we calculate a dilution factor of $D_0=0.13\pm0.02$ for NG 0522-2518 017220. 
We further compute the centre of flux of the third light in the aperture to be at $\vec{x}_\mathrm{backg. obj.}=(1.9\pm0.2, 1.7\pm0.2)$ pixel.
This information on $D_0$ and $\vec{x}_\mathrm{backg. obj.}$ is used as Gaussian priors on these parameters in our MCMC model fit.  

The object shows significant out-of-eclipse (OOE) modulation on a ${\sim}1$~per-cent level, which appears to be sinusoidal and in phase with the eclipse signal. 
We first analyse the time evolution of the OOE variation by dividing the lightcurve in equal sections in time, and compare the variability between these sections. We find that the OOE modulation significantly changes over the $175$ day observing span.
We further compute Lomb Scargle periodagrams for the OOE data. We identify multiple periods, which we can relate to the orbital period of the system and systematics on a sidereal day period.
With an orbital period of $\sim1.67$ days, it can be assumed that the orbit is circular and that the binary components are tidally locked.
Hence, we conclude that the OOE modulations may result from a combination of 1) stellar spots on either or both bodies, 2) a difference in the reflection indices of the two bodies, and 3) systematics introduced by the third light in the aperture. 

We consequently remove the OOE variation from the flux and centroid time series using Gaussian Process Regression, before analysing the time series with our MCMC model. We employ a combination of a Matern 3/2 kernel function, a linear kernel and a white noise kernel to model the OOE data of each time series individually. We then extrapolate the model and evaluate it for the eclipse data, resulting in the detrended lightcurve and centroid curves shown  Figs.~\ref{fig:CENTD_identification_NG0522-2518_017220_TEST18}A and \ref{fig:NG0522-2518_017220_vs_017221}.

The results of our MCMC model fit are shown in Fig.~\ref{fig:CENTD_identification_NG0522-2518_017220_TEST18} and Tab.~\ref{tab:Results}. Fig.~\ref{fig:NG0522-2518_017220_TEST18_centroid_fit CORNER} shows the resulting posterior distributions. 
We find a high inclination ($i = {80.14^{\circ}}\pm0.34^{\circ}$) of the system and comparable surface brightness ratio ($J=0.722\pm0.010$) of the two components. The radius ratio of the two bodies is estimated to be $R_2/R_1 = 0.2414\pm0.0044$.

The example of NG 0522-2518 017220 illustrates how the centroiding technique can aid us in identifying which system in the aperture shows the eclipsing signal. In particular, any visual information on the blend can be input as priors to refine the model fit and establish the physical parameters of the eclipsing system.

\begin{table}
  \centering
  \caption{Statistical identification of a centroid shift correlated to the transit-like signals in NG 0522-2518 017220 and NG 0409-1941 020057.
The table displays the signal-to-noise ratio (SNR) of the rolling correlation and cross-correlation analyses, displayed in Fig.~\ref{fig:CENTD_identification_NG0522-2518_017220_TEST18}
and \ref{fig:CENTD_identification_NG0409-1941_020057_TEST18}. 
Further it lists the resulting p-values from a T-test and binomial test of the in-transit centroid data, testing the Null Hypothesis that the centroid is distributed around the mean of the out-of-transit data, i.e. around $0$.}
    \begin{tabular}{l | rr}
          \hline
          \hline
          & \multicolumn{1}{r}{x} & \multicolumn{1}{r}{y} \\
          \hline
           \multicolumn{3}{c}{\textit{NG0522-2518 017220}} \\
   			  SNR roll. corr. & 41.43  & 43.01 \\
  			  SNR cross-corr. & 53.28  & 26.55 \\
  			  p-value T-test & $4.01\cdot10^{-32}$ & $5.69\cdot10^{-26}$ \\
  			  p-value Binomial test & $6.62\cdot10^{-24}$ & $9.96\cdot10^{-18}$ \\
  			  \hline
           \multicolumn{3}{c}{\textit{NG0409-1941 020057}} \\
 			   SNR roll. corr. & 51.66  & 44.06 \\
 			   SNR cross-corr. & 15.75 & 12.28 \\
			    p-value T-test & $7.92\cdot10^{-26}$ & $4.34\cdot10^{-16}$ \\
			    p-value Binomial test & $8.67\cdot10^{-19}$ & $5.65\cdot10^{-12}$ \\
    \end{tabular}%
  \label{tab:NG 0522-2518 017220}%
\end{table}%

\subsubsection{NG 0409-1941 020057}
\label{sss:NG 0409-1941 020057}

\begin{figure}
 \includegraphics[width=\columnwidth]{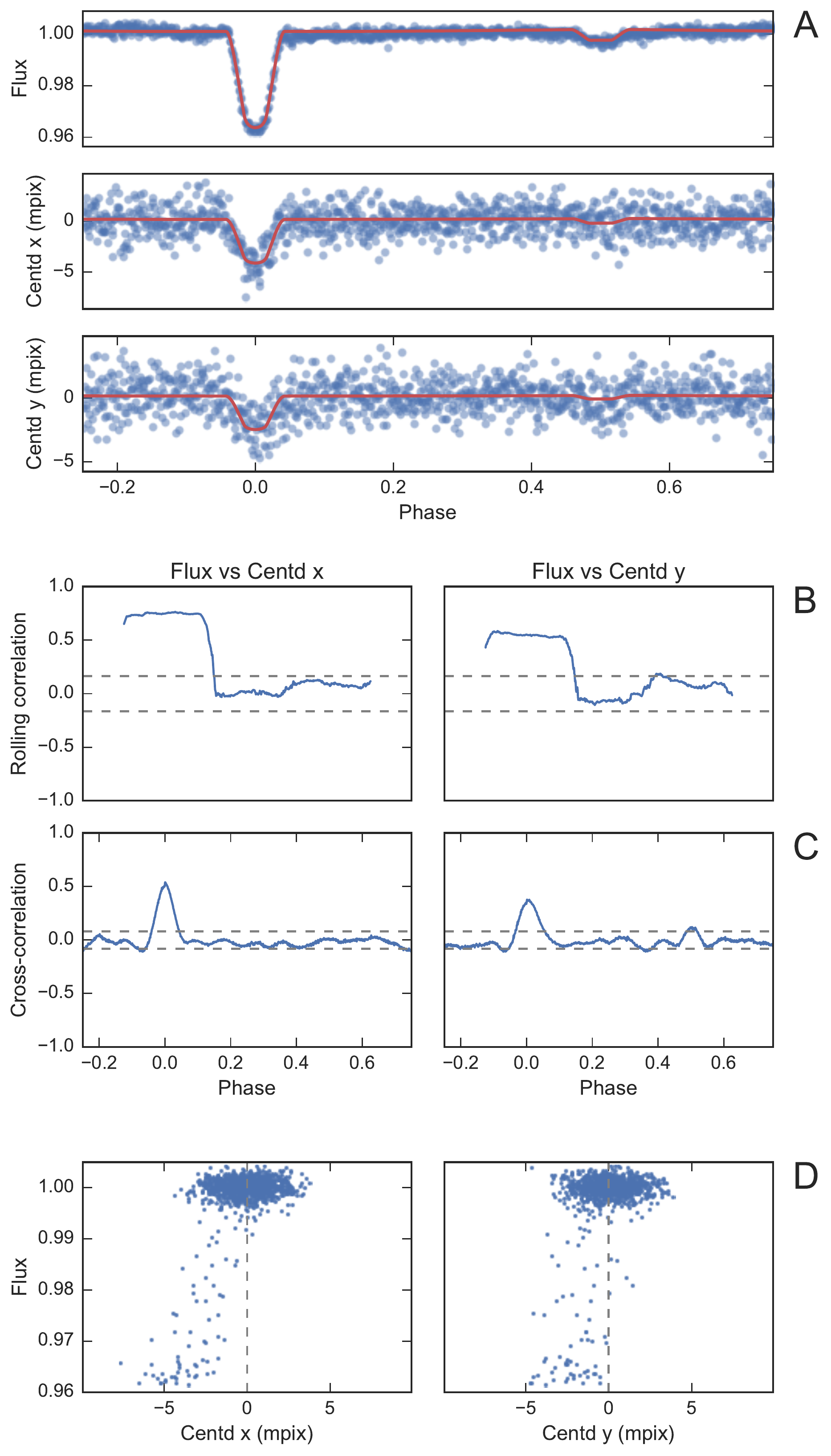}
 \caption{Identification and model fit of a centroid shift correlated to the transit signal in NG 0409-1941 020057. 
 See caption of Fig.~\ref{fig:CENTD_identification_NG0522-2518_017220_TEST18}.
 }
 \label{fig:CENTD_identification_NG0409-1941_020057_TEST18}
\end{figure}

The transit-like signal of NG 0409-1941 020057 was initially detected with a period of ${\sim}0.8$~days and width of ${\sim}3.4$~hours. However, with additional \textit{NGTS} photometry the true period could be established as ${\sim}1.61$~days, with a ${\sim}4$~per-cent primary eclipse and a ${\sim}0.5$~per-cent secondary eclipse. The object is well suited for follow-up at $G=13.3$, $J=12.1$ and $K=11.7$. It is identified as \mbox{2MASS 04104778-2031575} and \mbox{GAIA 5091012688012721664}, and is located at \mbox{RA = 04h 10m 47.8s}, \mbox{DEC = -20d 31m 57.5s}. 
The signal, however, is significantly correlated with ${\sim}5$~mpix and ${\sim}2$~mpix centroid shifts in x and y, respectively (see Fig.~\ref{fig:CENTD_identification_NG0409-1941_020057_TEST18}, Tab.~\ref{tab:NG 0522-2518 017220}).

The source is listed as a single source in various catalogues, including \textit{2MASS} and \textit{GAIA} DR1. 
We investigate the archival images, but find no definite indication of two seperate sources, despite a marginal ellipticity of the \textit{2MASS} point-spread-function.
The latest release of \textit{GAIA} DR1, however, is incomplete below ${\sim}4$~arcsec separation\footnote{see e.g. \url{https://www.cosmos.esa.int/web/gaia/dr1} (online 12 May 2017) }.
We employ this incompleteness in our MCMC model as an upper limit for uniform priors on the relative CCD position of the background object.
Due to the short orbital period and the clear secondary eclipse at phase $0.5$, we restrict our MCMC model to circular orbits. 

The results of our MCMC analysis can be seen in Figs.~\ref{fig:CENTD_identification_NG0409-1941_020057_TEST18} and \ref{fig:NG0409-1941_020057_TEST18_centroid_fit CORNER}, and are summarised in Tab.~\ref{tab:Results}.
The object undergoing the eclipse is highly diluted with $\mathrm{D_0}=0.849^{+0.010}_{-0.015}$, indicating that the signal originates from a blended background object. 
This background object would hence show an undiluted transit depth of $\delta_e = 24.5^{+2.0}_{-2.6}$~per-cent, and has a surface brightness ratio of $J=0.1061\pm0.0065$ and radius ratio of $0.462^{+0.018}_{-0.022}$.

\begin{figure}
 \subfigure{\includegraphics[width=\columnwidth]{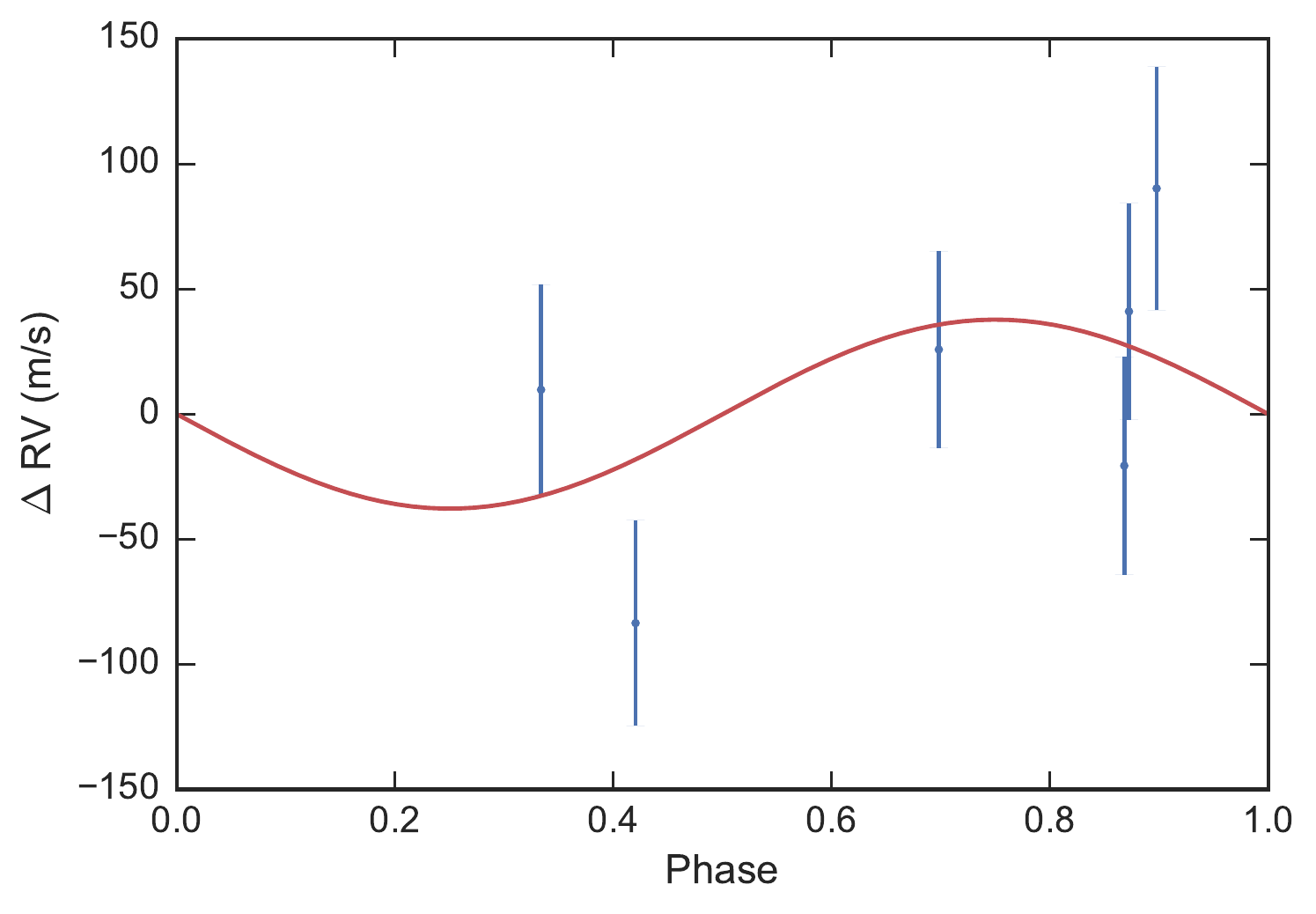}}
 \subfigure{\includegraphics[width=\columnwidth]{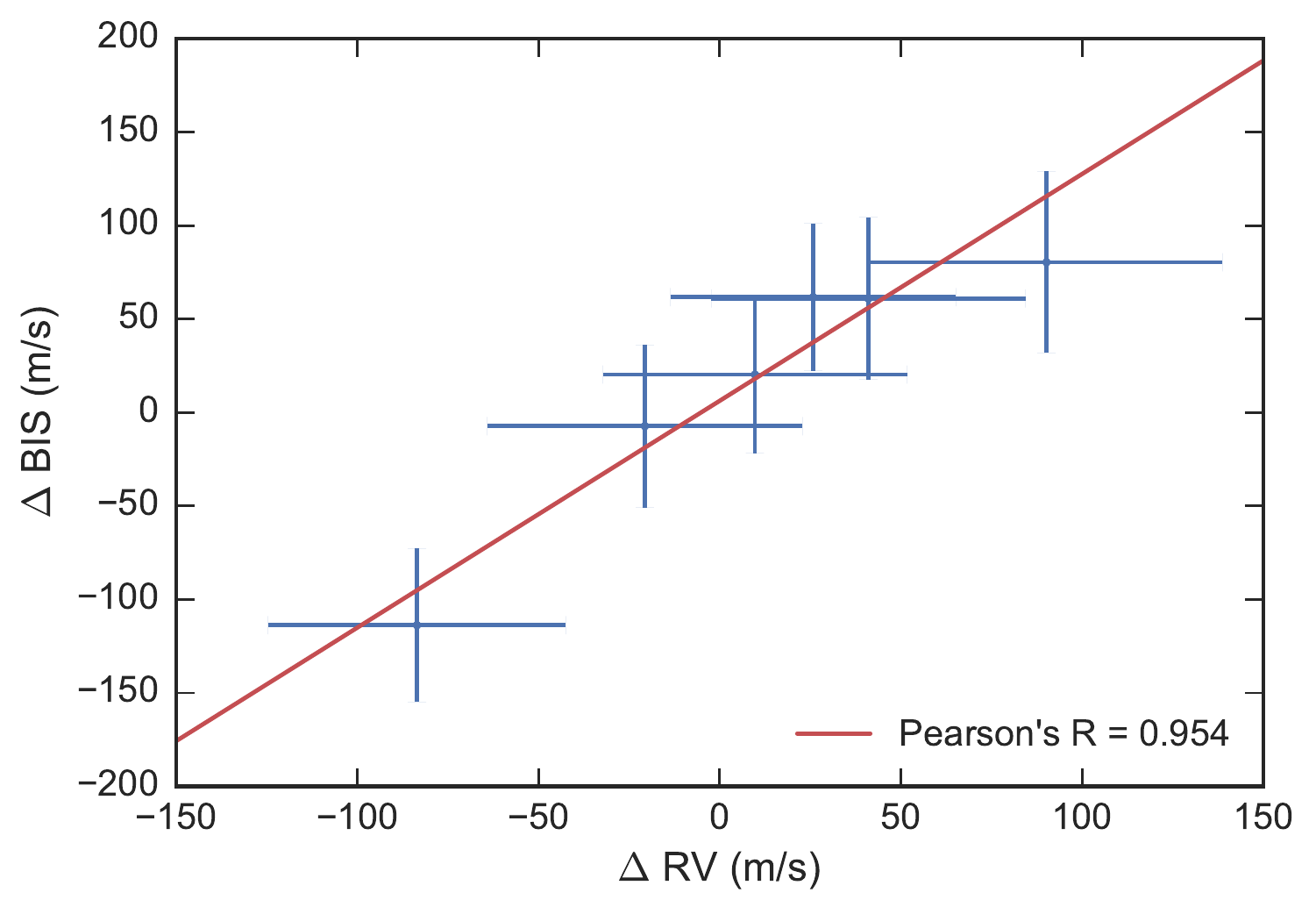}}
 \caption{\textit{Coralie} radial velocity (RV) measurements of NG 0409-1941 020057. Upper panel: RV signal phase folded on the photometric period and epoch. The solid line indicates the best fit Keplerian solution assuming a circular orbit and the photometric period and epoch. Lower panel: RV bisector spans against the measured RV signal. The strong correlation (Pearson's R = $0.954$) indicates the detected RV variations are due to a diluted spectrum shifting at large amplitudes, such as a background eclipsing binary.}
 \label{fig:RV_NG0409-1941_020057}
\end{figure}

Before the correct period had been established and our centroiding analysis had been performed, six reconnaissance radial velocity (RV) measurements were taken using the \textit{Coralie} spectrograph \citep{Queloz2001} on the Swiss 1.2\,m telescope at La Silla Observatory, Chile. These measurement are set out in Table~\ref{tab:RV}. The RV signal for this system, when phase-folded at the true period, shows an in-phase variation of approximately 50\,m\,s$^{-1}$ (see Fig~\ref{fig:RV_NG0409-1941_020057}). Such a signal is consistent with what may be expected for a typical hot Jupiter. However the bisectors of the RV cross-correlation functions show a significant correlation with the RV amplitude (see Fig.~\ref{fig:RV_NG0409-1941_020057}) This indicates that the variation seen for this target is due to a blended star that is spectroscopically contaminating the cross-correlation function and is moving in phase with the photometric period.

Note that the acceptance of the \textit{Coralie} fibre is ${\sim}2$~arcsec. The results from our MCMC model predict the blended background object to be offset from the target by $0.653\pm0.040$~arcsec in x and $0.396\pm0.034$~arcsec in y, further supporting the hypothesis that the \textit{Coralie} signal is affected by the blend. The evidence provided by both the centroid vetting and the RV data are in agreement and suggest a highly diluted background eclipsing binary (BEB) scenario. \\

\begin{table}
  \centering
  \caption{\textit{Coralie} radial velocities of NG 0409-1941 020057}
      \begin{tabular}{lcccc}
      \hline
      \hline
BJD & RV & RV error & BIS\\
(-2400000)& (km\,s$^{-1}$)&(km\,s$^{-1}$) & (km\,s$^{-1}$)\\
      \hline
57605.906507&	103.93259&	0.04857&	 0.08040\\
57613.901203&	103.88342&	0.04335&	 0.06101\\
57630.852688&	103.75888&	0.04108&	-0.11377\\
57632.906217&	103.86820&	0.03938&	 0.06171\\
57634.786375&	103.82177&	0.04350&	-0.00728\\
57638.748983&	103.85213&	0.04198&	0.02030\\
    \end{tabular}%
  \label{tab:RV}%
\end{table}%


\section{Discussion}
\label{s:Discussion}

\subsection{Impact of the centroiding technique on \textit{NGTS} candidate vetting}
\label{ss:Impact of the centroiding technique on NGTS candidate vetting}

\textit{NGTS} is the first ground-based wide-field transit survey to employ the centroid technique for automated and routine candidate vetting. 
The presented algorithm is already part of the \textit{NGTS} pipeline employed for all detected transit-like signals.
We achieve an average centroid precision of $0.75$~milli-pixel for all candidates, and as low as $0.25$~milli-pixel for individual objects.
This precision depends on the photometric data quality as a direct consequence of Eq.~\ref{eq:BEB} and \ref{eq:BCS}. We therefore expect to observe an increase of the centroiding utility with higher signal-to-noise ratios of the detected transit-like signal.

\begin{figure}
 \includegraphics[width=\columnwidth]{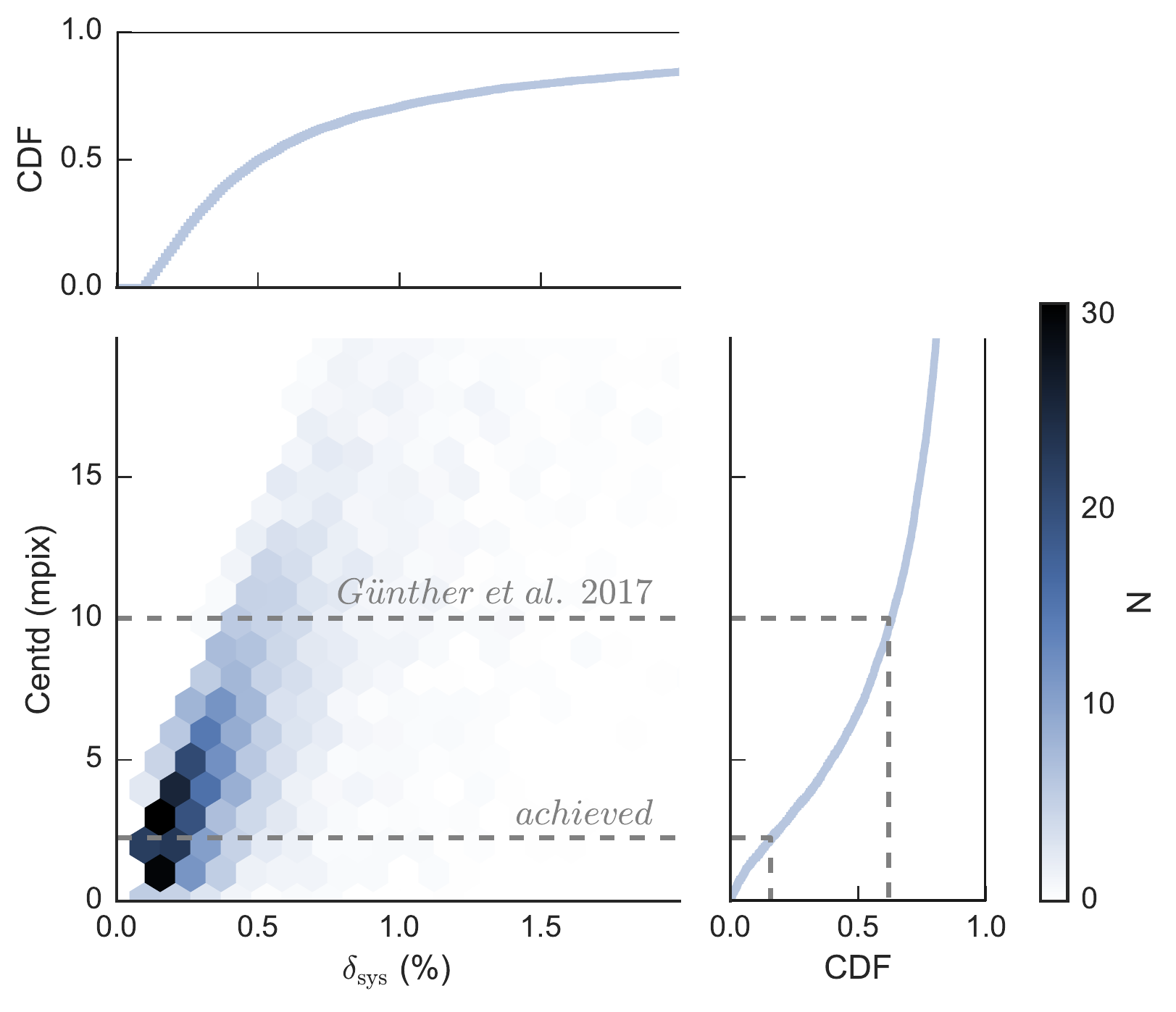}
 \caption{The impact of the centroiding algorithm for automated planet candidate vetting in \textit{NGTS}, based on the simulated yield for four years of survey operation. Shown is the total number of BEBs triggering a signal as a function of the measured (diluted) transit depth of the system, $\delta_\mathrm{sys}$, and the caused centroid shift, $\Delta \xi$, as well as the cumulative distribution function (CDF) of each parameter. Assuming that all centroid signals $>3\sigma$ above the noise level of $0.75$~mpix can be detected, the achieved centroiding detection threshold exceeds our previous assumptions by an order of magnitude. This will allow to directly identify $\sim80$~per-cent of all BEBs.
}
 \label{fig:Impact_of_centroiding}
\end{figure}

The obtained centroid precision exceeds the previous assumptions in our yield estimations by an order of magnitude \citep{Guenther2017}.
The yield simulator is based on a galactic model and the planet occurrence rates estimated by the Kepler mission. It considers the measured noise levels and observation window function of \textit{NGTS}, and simulates vetting criteria to identify false positives. 
We previously assumed all centroid shifts $>10$~mpix could be detected, and predicted that $\sim38$~per-cent of all background eclipsing binaries (BEBs) can be identified with \textit{NGTS} alone.
Considering the achieved centroid precision of $0.75$~mpix, we update our yield simulator and repeat this analysis.
We assume that all centroid signals $>3\sigma$ above the noise level can be detected. This is motivated by the detection of a $\sim2$~mpix centroid shift for NG 0409-1941 020057 (section~\ref{sss:NG 0409-1941 020057} Fig.~\ref{fig:CENTD_identification_NG0409-1941_020057_TEST18}). We estimate that this allows to directly identify $\sim80$~per-cent of all BEBs (see Fig.~\ref{fig:Impact_of_centroiding}).

To verify these estimations we test our implementation on eclipsing systems from the latest \textit{NGTS} pipeline run. 
We restrict this comparison to a sample of obvious astrophysical signals with eclipse depths $>2$~per-cent to avoid the influence of spurious signals. Note that this sample is mostly comprised of undiluted eclipsing binaries due to their high occurrence rates.
We find that $16\pm8$~per-cent of this sample show a significant correlation between their photometric flux and centroid data. The given confidence interval is the standard error of the mean of a sample of binomial random variables.
In comparison, from our simulations we would expect to identify a centroid shift for $12\pm2$~per-cent of this sample.
These findings are consistent and highlight the expected success of the centroid algorithm for the automated candidate vetting pipeline.

\subsection{Investigated candidates}
\label{ss:Investigated candidates}

In sections \ref{sss:NG 0522-2518 017220} and \ref{sss:NG 0409-1941 020057} we demonstrated our method in two case studies. 
First, visual information on NG 0522-2518 017220 allowed us to identify which blended object is undergoing the eclipse. We identify out-of-eclipse modulation, which is likely due to either a) star spots, b) the reflection effect, or c) systematics from the blended contaminant, or a combination of these effects. The eclipsing system is shown to be a grazing low-mass binary, likely consisting of a $K$ star primary and $M$ star or brown dwarf secondary.

Second, an analysis of NG 0409-1941 020057 reveals that its signal in fact originates from a highly diluted source, and thus suggests a deep background eclipsing binary (undiluted depth of $24.5^{+2.0}_{-2.6}$~per-cent) as a cause. 
RV data previously collected with \textit{Coralie} shows a correlation in the bisectors of the RV cross-correlation function, which supports the results of our centroiding method.

For the latter candidate, the blended objects are estimated to be $<1$~arcsec separated, such that no existing catalogue has resolved the system. This demonstrates the effectiveness of the centroid technique, as it predicts the relative position of blended background objects, which will eventually be confirmable with the resolution of upcoming results from \textit{GAIA}.

The parameter space was thoroughly explored using differential evolution algorithms and multiple MCMC runs from different starting positions. Additionally, care was taken to reach convergence of all walkers and to not fall into local minima. However, systematics in the flux and centroid time series may still be present after the detrending procedure, and could potentially restrict the exploration of the parameter space. This could lead to the underestimation of MCMC posterior likelihood distributions. Hence, a future refinement of the presented work could be the implementation of a joint MCMC model directly incorporating Gaussian Process Regression \citep[see e.g.][]{Pepper2017, Gillen2017}.


\section{Conclusion}
\label{s:Conclusion}
We developed a comprehensive framework to extract and detrend flux centroid information from \textit{NGTS} data. The introduced algorithms are part of an automated vetting pipeline for all \textit{NGTS} candidates. 
We achieve an average precision of $0.75$~milli-pixel on the phase-folded centroids over an entire field, and $0.25$~milli-pixel for specific targets.
This enables the identification of systems that are currently too close ($< 4$ arcsec) to be resolved with any photometric or astrometric all-sky survey.
Case studies of \textit{NGTS} candidates illustrate that different scenarios can lead to a centroid motion, yet our robust MCMC fitting procedure is able to determine the true origin of a given transit-like signal.
In total, we estimate to be able to rule out $\sim$80 per-cent of all blended variable background objects with \textit{NGTS} data alone. These systems would otherwise have to be followed-up with higher-resolution imaging, or may otherwise potentially be miss-identified as planet candidates.
While the centroiding technique has previously been employed for the space-based \textit{Kepler} mission, this is the first time it has been implemented in a ground-based wide-field photometric survey.


\section*{Acknowledgements}
\addcontentsline{toc}{section}{Acknowledgements}
This research is based on data collected under the \textit{NGTS} project at the ESO La Silla Paranal Observatory.
\textit{NGTS} is operated with support from an UK Science and Technology Facilities Council (STFC) research grant (ST/M001962/1). 
This work has further made use of data from the European Space Agency (ESA) mission {\it Gaia} (\url{https://www.cosmos.esa.int/gaia}), processed by
the {\it Gaia} Data Processing and Analysis Consortium (DPAC, \url{https://www.cosmos.esa.int/web/gaia/dpac/consortium}). Funding for the DPAC has been provided by national institutions, in particular
the institutions participating in the {\it Gaia} Multilateral Agreement. 
Moreover, this publication makes use of data products from the Two Micron All Sky Survey, which is a joint project of the University of Massachusetts and the Infrared Processing and Analysis Center/California Institute of Technology, funded by the National Aeronautics and Space Administration and the National Science Foundation.
We also make use of the open-source Python packages {\scshape numpy} \citep{vanderWalt2011}, {\scshape scipy} \citep{Jones2001}, {\scshape matplotlib} \citep{Hunter2007}, {\scshape pandas} \citep{McKinney2010}, {\scshape emcee} \citep{Foreman-Mackey2013}, {\scshape corner} \citep{Foreman-Mackey2016}, and {\scshape eb} \citep{Irwin2011}.
The latter is based on the previous {\scshape JKTEBOP} \citep{Southworth2004a, Southworth2004b} and {\scshape EBOP} codes \citep{Popper1981}, and models by \cite{Etzel1981}, \cite{Mandel2002}, \cite{Binnendijk1974a,Binnendijk1974b}, and \cite{Milne1926}.
DJA is funded under STFC consolidated grant reference ST/P000495/1.
MNG is supported by the UK Science and Technology Facilities Council (STFC) award reference 1490409 as well as the Isaac Newton Studentship.


\bibliographystyle{mnras}
\bibliography{Guenther2017_References}


\appendix

\section*{Appendix}
\renewcommand{\thefigure}{A\arabic{figure}}
\renewcommand{\thetable}{A\arabic{table}}
\renewcommand{\thesubsection}{\Alph{subsection}}

\begin{table*}
  \caption{Parameters of the blended eclisping systems NG 0522-2518 017220 and NG 0409-1941 020057.}
\scalebox{1}{
    \begin{tabular}{llcc}
    \hline
    \hline
          &       & NG 0522-2518 017220 & NG 0409-1941 020057 \\
    \hline
    \textit{Catalogue values} &       &       &         \\
    \hline
	Coordinates &       & \mbox{RA = 05h 23m 31.6s} & \mbox{RA = 04h 10m 47.8s}   \\
      &       & \mbox{DEC = -25d 08m 48.4s} & \mbox{DEC = -20d 31m 57.5s} \\
    2MASS ID &       & \mbox{2MASS 05233161-2508484} & \mbox{2MASS 04104778-2031575}   \\
    \textit{Gaia} ID &       & \mbox{GAIA 2957881682551005056} & \mbox{GAIA 5091012688012721664}   \\
    Magnitudes &       & $G=13.6$, $J=12.6$, $K=12.2$ & $G=13.3$, $J=12.1$ and $K=11.7$   \\
    Colour &       & $J-K = 0.4$ & $J-K = 0.4$  \\
    \hline
    \textit{Fitted parameters} &       &       &         \\
    \hline
    $\Delta x$ & Relative CCD x position of the blend in pixel & $1.67\pm0.14$ & $0.1312\pm0.0071$   \\
    $\Delta y$ & Relative CCD y position of the blend in pixel & $1.87\pm0.14$ & $0.0798\pm0.0064$   \\
    $F_0$ & Offset in normalised flux & $(7.2\pm1.1)\cdot 10^{-4}$ & $(9.68\pm0.50)\cdot 10^{-4}$  \\
    $\xi_{x,0}$ & Offset in centroid in x in pixel & $(-2.7\pm5.8) \cdot 10^{-5}$ & $(1.93\pm0.47) \cdot 10^{-4}$   \\
    $\xi_{y,0}$ & Offset in centroid in y in pixel & $(-6.0\pm5.9)\cdot 10^{-5}$ & $(1.61\pm0.44)\cdot 10^{-4}$   \\
    $D$   & Dilution & $0.1217^{+0.0099}_{-0.0088}$ & $0.849^{+0.010}_{-0.015}$   \\
    $J$   & surface brightness ratio & $0.722\pm0.010$ & $0.1061\pm0.0065$   \\
    $(R_1+R_2)/a$ & Sum of radii over semi-major axis & $0.2414\pm0.0044$ & $0.2634\pm0.0031$   \\
    $R_2/R_1$ & Ratio of radii & $0.247\pm0.012$ & $0.462^{+0.018}_{-0.022}$   \\
    $\cos{i}$ & Cosine of the inclination & $0.1713\pm0.0051$ & $0.022^{+0.021}_{-0.015}$   \\
    $P$   & Period in seconds & $140145.80\pm0.59$ & $138852.0\pm1.4$   \\
    $T_0$ & Epoch in seconds & $54443339\pm35$ & $59237250\pm34$   \\
    $e$   & Eccentricity & $0.0$ (fixed) & $0.0$ (fixed)   \\
    $\omega$ & Argument of periastron in degree & $0.0$ (fixed) & $0.0$ (fixed)   \\
    $\sigma(F)$ & Error on normalised flux & $(7.805\pm0.037)\cdot 10^{-3}$ & $(4.835\pm0.032\cdot 10^{-3}$   \\
    $\sigma(\xi_x)$ & Error on centroid in x & $(8.383\pm0.041)\cdot 10^{-3}$ & $(4.787\pm0.032)\cdot 10^{-3}$   \\
    $\sigma(\xi_y)$ & Error on centroid in y & $(8.468\pm0.041)\cdot 10^{-3}$ & $(4.448\pm0.03)\cdot 10^{-3}$   \\
    \hline
    \textit{Derived parameters} &       &       &         \\
    \hline
    $i$   & Inclination in degree & $80.14\pm0.34$ & $88.61^{+0.86}_{-1.26}$   \\
    $R_1/a$ & Radius of the primary over semi-major axis & $0.1935\pm0.0027$ & $0.1800^{+0.0037}_{-0.0020}$   \\
    $R_2/a$ & Radius of the secondary over semi-major axis & $0.0478^{+0.0030}_{-0.0027}$ & $0.0830^{+0.0030}_{-0.0033}$   \\
    $T_\mathrm{2}$ & Midpoint of secondary eclipse in s & $54513413\pm41$ & $59306676\pm36$   \\
    $T_\mathrm{dur,1}$ & Duration of primary eclipse in s & $185700\pm2100$ & $281100\pm3600$   \\
    $T_\mathrm{dur,2}$ & Duration of secondary eclipse in s & $185700\pm2100$ & $281100\pm3600$   \\
    $\delta_{1,\mathrm{dil}}$ & Diluted depth of the primary eclipse & $3.252^{+0.043}_{-0.051}$ & $3.628^{+0.032}_{-0.047}$   \\
    $\delta_{2,\mathrm{dil}}$ & Diluted depth of the secondary eclipse & $2.798^{+0.041}_{-0.052}$ & $0.238\pm0.023$   \\
    $\delta_{1,\mathrm{undil}}$ & Undiluted depth of the primary eclipse & $3.716^{+0.067}_{-0.073}$ & $24.5^{+2.0}_{-2.6}$   \\
    $\delta_{2,\mathrm{undil}}$ & Undiluted depth of the secondary eclipse & $3.197^{+0.063}_{-0.070}$ & $2.10\pm0.26$   \\
    $\Delta x_\mathrm{sky}$ & Relative sky position of the blend in arcsec & $8.29\pm0.76$ & $0.653\pm0.040$   \\
    $\Delta y_\mathrm{sky}$ & Relative sky position of the blend in arcsec & $9.29\pm0.84$ & $0.396\pm0.034$  \\
    \end{tabular}%
    }
  \label{tab:Results}%
\end{table*}%

\begin{figure*}
 \includegraphics[width=2\columnwidth]{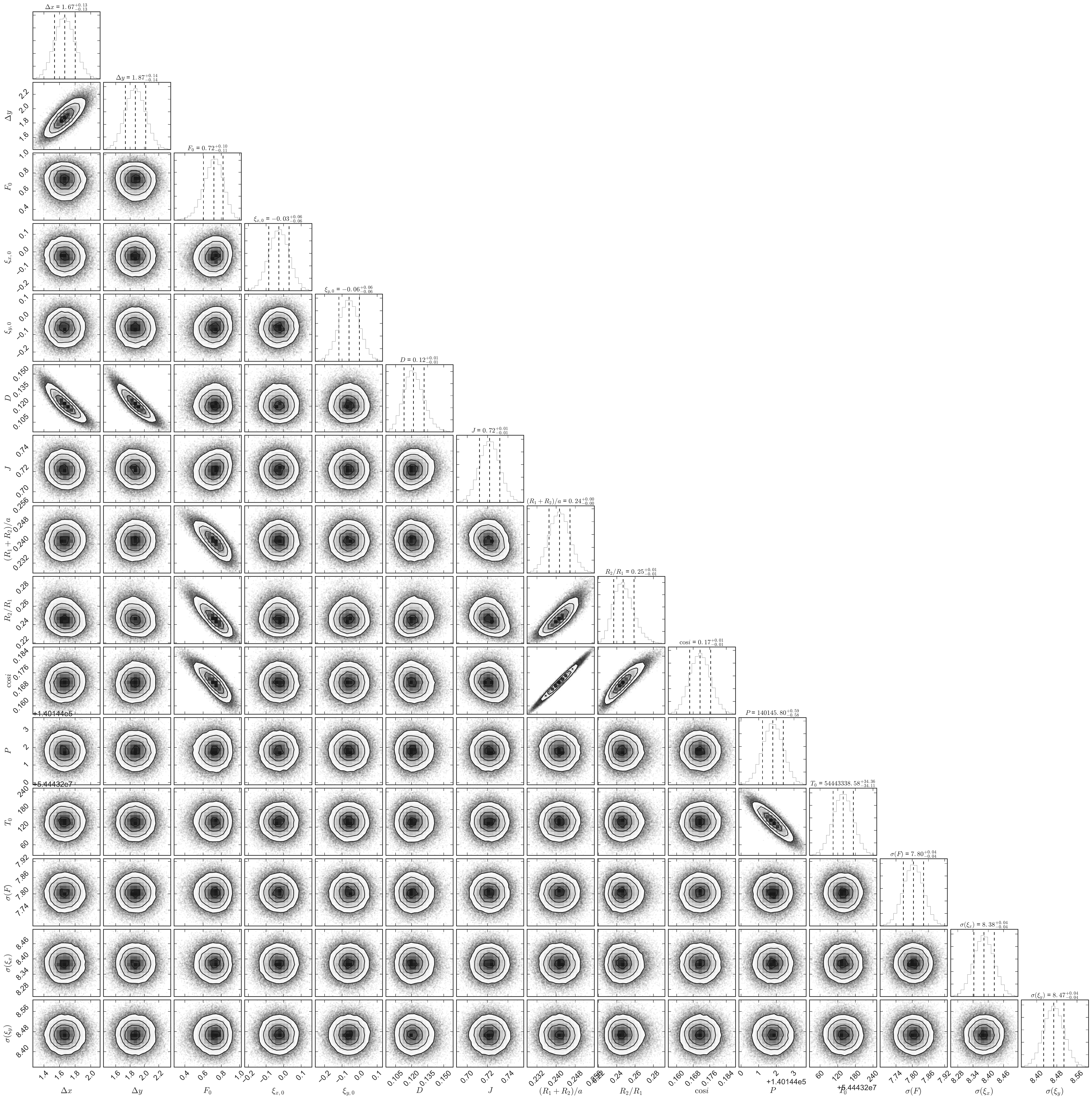}
 \caption{Posterior likelihood distributions for all parameters of the MCMC fit to NG0522-2518 017220. A description of the model and all parameters can be found in section~\ref{sss:Bayesian analysis}.}
 \label{fig:NG0522-2518_017220_TEST18_centroid_fit CORNER}
\end{figure*}

\begin{figure*}
 \includegraphics[width=2\columnwidth]{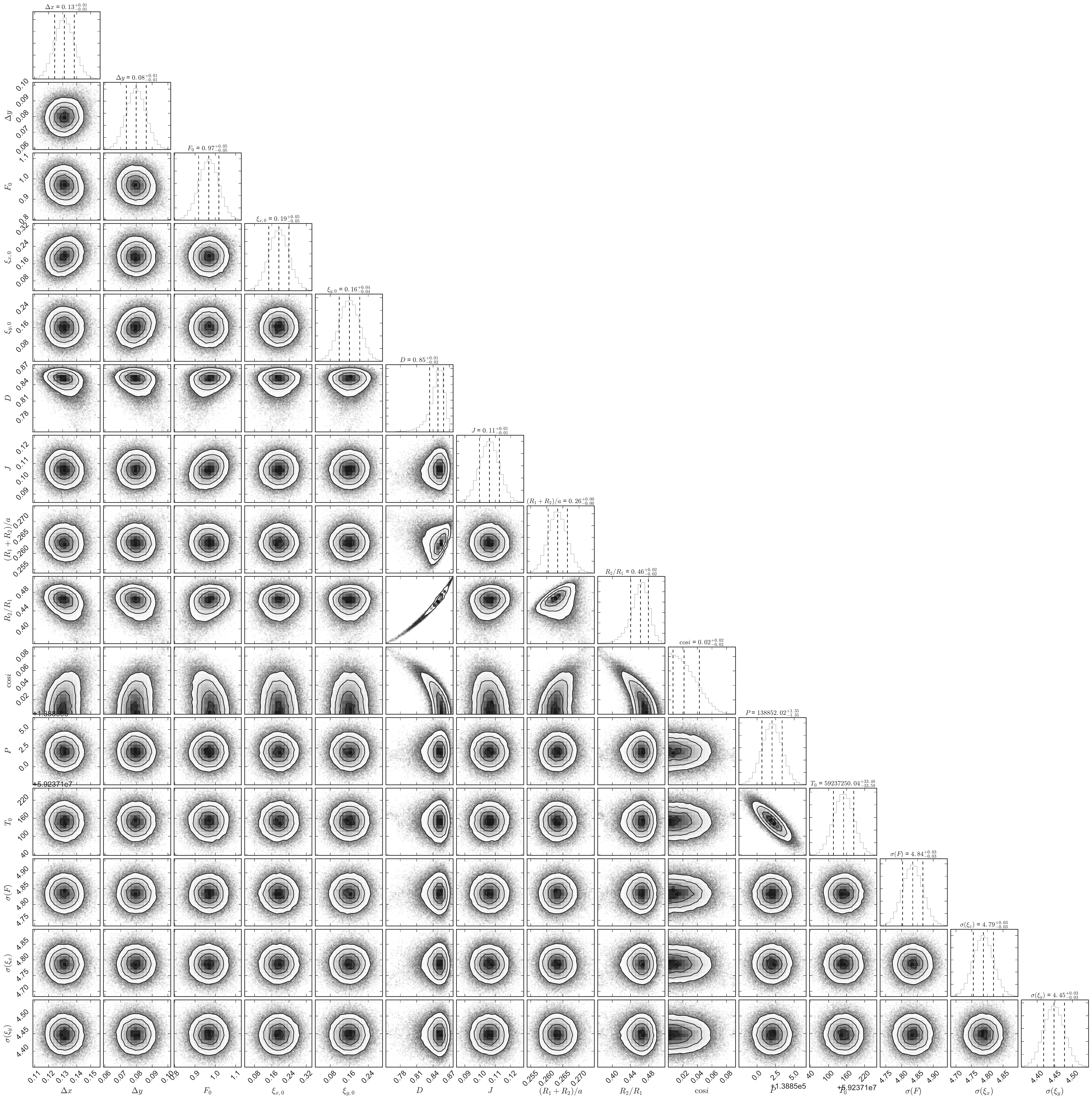}
 \caption{Posterior likelihood distributions for all parameters of the MCMC fit to NG0409-1941 020057. A description of the model and all parameters can be found in section~\ref{sss:Bayesian analysis}.}
 \label{fig:NG0409-1941_020057_TEST18_centroid_fit CORNER}
\end{figure*}


\bsp	
\label{lastpage}
\end{document}